\documentclass{achemso}

\usepackage[version=3]{mhchem}
\usepackage{siunitx}
\usepackage{booktabs}
\usepackage[hidelinks]{hyperref}
\usepackage{amssymb}
\usepackage{bm}
\usepackage{mathtools}

\DeclareSIUnit\angstrom{\text {Å}}
\DeclareSIUnit[quantity-product = ]\percent{\char`\%}

\author{Yuchen Lou}
\affiliation{Department of Chemistry, Molecular Sciences Research Hub, White City Campus, Imperial College London, Wood Lane, London W12 0BZ, UK}

\author{Alex M. Ganose}
\affiliation{Department of Chemistry, Molecular Sciences Research Hub, White City Campus, Imperial College London, Wood Lane, London W12 0BZ, UK}
\email{a.ganose@imperial.ac.uk}

\title{Discovery of highly anisotropic dielectric crystals with equivariant graph neural networks}

\begin{document}

\begin{abstract}
Anisotropy in crystals plays a pivotal role in many technological applications. For example, anisotropic electronic and thermal transport are thought to be beneficial for thermoelectric applications, while anisotropic mechanical properties are of interest for emerging metamaterials, and anisotropic dielectric materials have been suggested as a novel platform for dark matter detection. Understanding and tailoring anisotropy in crystals is therefore essential for the design of next-generation functional materials. To date, however, most data-driven approaches have focused on the prediction of scalar crystal properties, such as the spherically averaged dielectric tensor or the bulk and shear elastic moduli. Here, we adopt the latest approaches in equivariant graph neural networks to develop a model that can predict the full dielectric tensor of crystals. Our model, trained on the Materials Project dataset of c.a.~6,700 dielectric tensors, achieves state-of-the-art accuracy in scalar dielectric prediction in addition to capturing the directional response. We showcase the performance of the model by discovering crystals with almost isotropic connectivity but highly anisotropic dielectric tensors, thereby broadening our knowledge of the structure-property relationships in dielectric crystals.
\end{abstract}

\section{Introduction}
Anisotropy refers to the directional non-uniformity of structures and physical properties and is especially relevant when considering tensorial properties such as elastic and dielectric tensors.
Anisotropic materials can offer a tailored directional response to external stimuli that isotropic materials cannot, making them indispensable in fields like electronics, energy, and quantum computing\cite{ohno_window_2010, oberg_control_2014, li_demonstration_2021}.
Recently, the demand for novel materials that meet specific performance criteria has surged.\cite{li_anisotropic_2023}
For example, barium titanate (\ce{BaTiO3}) has been the main component of ceramic capacitors for over 70 years, owing to its high dielectric constant at room temperature and chemical stability.\cite{pan_brief_2010} 
However, in recent years, these traditional materials no longer suffice due to the evolving needs of the electronics industry\cite{sun_review_2022}.
Notably, advancements in dynamic random access memory (DRAM) and light-emitting diodes (LEDs) require additional properties such as unique anisotropy\cite{li_review_2021}.
Another area of interest for anisotropic dielectric properties is in birefringent crystals which have found use in display technologies and medical diagnostics\cite{palmer_x-ray_2014}.
At the same time, novel use cases for anisotropic materials have been proposed, including as dark matter detectors\cite{coskuner_directional_2021,griffin_multichannel_2020}.
As a result, there is an increasing demand to discover new anisotropic dielectrics.

When an electric field is applied to a material, the charges separate and create a net dipole moment that shields the electric field.
The strength of this shielding is quantified by the dielectric response.\cite{nurul_hidayah_thermoplastic_2021, spaldin_beginners_2012}.
When the frequency of the electric field is low, both ions and electrons can be displaced and respond to the oscillation of the field.\cite{izgorodina_components_2009} 
The total response from both charge carriers is termed the static dielectric constant, $\bm{\varepsilon}_\mathrm{s}$. 
When the frequency increases, the ions become too slow to respond to the oscillations and the ionic contribution to the dielectric response decreases.
The point at which the ions stop contributing and the response is solely due to the electrons is termed the high-frequency or optical dielectric constant, $\bm{\varepsilon}_\infty$.
Unlike the dielectric response of liquids and isotropic crystals which can typically be described by a scalar, the dielectric tensor characterises the full directional response of a crystal.

While, traditionally dielectric tensors are measured experimentally, this requires expensive and time-intensive synthesis and characterisation. 
Advancements in computing power and \textit{ab initio} simulations based on density-functional theory (DFT) have enabled the routine calculation of dielectric tensors.\cite{lejaeghere_reproducibility_2016}
In 2017, \citet{petousis_high-throughput_2017} used high-throughput density-functional perturbation theory (DFPT) to obtain the static and high-frequency dielectric tensors of 1,056 compounds, with the results added to the Materials Project database.\cite{jain_commentary_2013} 
Despite these advancements, the dielectric tensor dataset on the Materials Project only constitutes a small fraction of the entire dataset. At the time of writing, there are 7,277 dielectric tensors and more than 80,000 non-metallic materials.
A primary challenge is the computational expense of DFPT calculations, which can require hundreds of CPU hours per structure.
Furthermore, as DFT scales cubically with the number of atoms, the simulation of larger cells can quickly become impractical.
Consequently, building a complete database of dielectric tensors solely using DFT calculations remains untractable.\cite{tawfik_predicting_2020}

Machine learning in the physical sciences has gained substantial interest in recent years due to the availability of GPU resources and advances in computational algorithms.  Machine learning can establish complex structure-property relationships, make predictions of material properties, and generate novel crystal structures.\cite{butler_machine_2018, dawid_modern_nodate}
A key advantage of machine learning models is their ability to predict material properties orders of magnitude faster than conventional \textit{ab initio} calculations.\cite{tawfik_predicting_2020} Message-passing graph neural networks (MPGNNs) are a subset of deep learning that operate on graphs composed of nodes and edges.\cite{wu_comprehensive_2021,bronstein_geometric_2021} This capability is ideal for use in chemistry and materials science since the data structures encountered in these domains are typically point clouds which are incompatible with convolutional neural networks.\cite{reiser_graph_2022}
Graphs are a natural representation for crystals. For example, atoms can be represented as nodes and bonds as edges. The features associated with nodes can encode information such as atom type or formal charges, while the edge features can contain information on the bond types or degree of conjugation.\cite{zhou_graph_2020, fung_benchmarking_2021, xu_how_2019, xie_crystal_2018} Such representations capture the full geometric characteristics of materials and also remain intuitive and easily comprehensible for humans.\cite{zhou_atom2vec_2018}

One example in the MPGNN domain is equivariant graph neural networks.\cite{thomas_tensor_2018, batzner_e3-equivariant_2022, liao_equiformerv2_2024} These models utilise basis functions such as spherical harmonics to represent features in a way that is guaranteed to transform predictably upon transformation of the input structure.\cite{freeman_design_1991, cohen_steerable_2016, worrall_harmonic_2017} Equivariant models alleviate the need for expensive data augmentation, which involves applying many rotations or translations to the training set to ensure the model learns to recognise all orientations through brute force. 
This approach also does not guarantee that the model will be consistent when predicting on data with unseen transformations.
Equivariant models accurately interpret orientations within a unit cell regardless of its configuration, which ensures the output adheres to the symmetry rules\cite{schutt_schnet_2018, thomas_tensor_2018}.
Equivariant models have been proposed to predict tensorial properties such as elastic tensors\cite{wen_equivariant_2024}.
However, for dielectric properties, most studies to date have  predicted the scalar polycrystalline dielectric constant.\cite{chen_frequency-dependent_2020,morita_modeling_2020,takahashi_machine_2020}
There are significantly fewer works aimed at predicting the full dielectric tensor.
\citet{grisafi_symmetry-adapted_2018} employed Gaussian process regression with irreducible spherical tensor representations to predict the static dielectric constant of water.
More recently, \citet{falletta2024unified} developed an  approach to predict the dielectric response through automatic differentiation of the electric enthalpy with respect to atomic positions and the electric field, showcasing excellent performance on $\alpha$-\ce{SiO2}.
However, both studies rely on significant training data from \textit{ab intio} molecular dynamics simulations while the trained models are only applicable to single systems.

In this paper, we develop an equivariant message-passing graph neural network, AnisoNet, for predicting the dielectric tensors of crystals across the periodic table.
AnisoNet takes a periodic structure as input and predicts the full 3$\times$3 dielectric tensor as output. AnisoNet is equivariant to the transformation of the input structure, meaning the output transforms predictably upon rotation, translation and inversion.
Furthermore, equivariance enforces that the symmetry of the output tensor is consistent with the symmetry of the input.
To quantify the degree of anisotropy of dielectric crystals, we introduce a new metric termed the ansiotropy ratio, $a_\mathrm{r}$.
Our model achieves state-of-the-art performance in both the prediction of the scalar dielectric constant and direction-dependent response.
We apply AnisoNet to the Materials Project dataset to discover novel highly anisotropic dielectric crystals.
The top 137 materials are validated using high-throughput DFPT calculations, finding an average $a_\mathrm{r}$ over 3.4 times larger than that of the training dataset.
Lastly, we obtain the full frequency-dependent dielectric response of 5 candidates and discuss the structure-property relationships that drive anisotropy in the solid state.

\section{Methodology}

\subsection{Irreducible representation of the dielectric tensor}

The dielectric tensor is a 3$\times$3 symmetric tensor that fully describes the dielectric response of a material.
The diagonal terms represent the dielectric response along the three Cartesian axes, and the off-diagonal terms describe the coupling between the axes.
The symmetry of the dielectric tensor follows $\varepsilon_{ij} = \varepsilon_{ji}$ (in indicial notation, where $i, j \in {1, 2, 3}$).
As such, only 6 of the 9 components of $\bm{\varepsilon}$ are independent.\cite{prati_propagation_2003}
As the dielectric constant describes the relative permittivity (with the absolute permittivity given by $\kappa = \bm{\varepsilon} \times \varepsilon_0$, where $\varepsilon_0$ is the vacuum permittivity), the minimum value of the independent components along the diagonal is 1\cite{prati_propagation_2003}.
An exception is metamaterials which can possess negative dielectric constants due to non-linear effects\cite{jahani_all-dielectric_2016}.
However, in this work we are only concerned with dielectric properties calculated through the linear response formalism which cannot describe such higher-order phenomena\cite{xie_recent_2022}.
As the values of the dielectric tensor components depend on the choice of the coordinate system (\textit{i.e.}, the rotation of the crystal in Cartesian space), it can be difficult to build predictive models for tensorial properties.
This can be handled through harmonic decomposition, where the space of all dielectric tensors is factored into the direct sum of irreducible representations of SO(3).
As such, any dielectric tensor can be written in the form
\begin{equation}
\bm{\varepsilon} = h_1(\lambda) + h_2(\mathbf{S}),
\label{eq:decomposition}
\end{equation}
where $\lambda$ is a scalar, $\mathbf{S}$ is a second-rank symmetric traceless tensor, and the functions $h_1$ and $h_2$ are constant for all systems.
Crucially, each part in Eqn.~\ref{eq:decomposition} transforms predictability with respect to SO(3) operations (rotations), thereby enabling the development of machine learning models that leverage the properties of equivariance.

\begin{figure}[t]
\centering
\includegraphics[width=\textwidth]{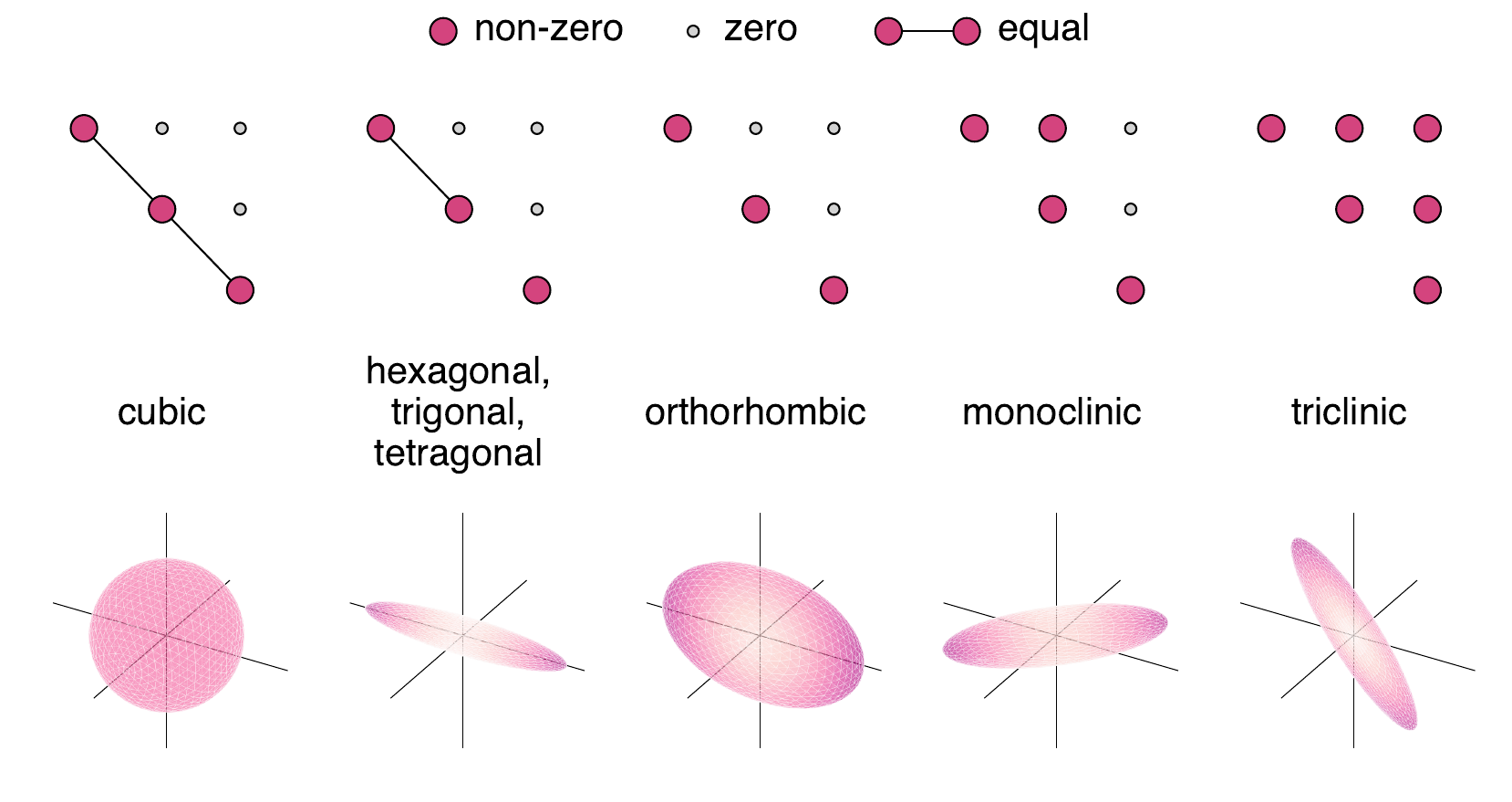}
\caption{Tensor and three-dimensional visualisation of dielectric tensors by crystal system. All tensors are symmetric about the main diagonal, with the lower triangular part omitted for clarity. Grey circles indicate zero components, pink circles indicate non-zero components, and pink circles connected by a line indicate the two components are dependent. The lower panels display the dielectric modulus surface in three dimensions.}
\label{fig:dielectric_tensor_form}
\end{figure}

The crystal symmetry will influence the form of the dielectric tensor. In cubic systems, the dielectric tensor assumes a diagonal form with only one independent component repeated along the diagonal. This can be visualised as a sphere in three dimensions, indicating the dielectric constant will be uniform in all directions it is measured. In tetragonal, hexagonal, and trigonal systems there are two independent components, with one of the diagonal components different from the other two. This corresponds to an obloid in three dimensions with the principle components oriented along the Cartesian axes. For orthorhombic systems, the tensor has three independent components on the diagonal with no off-diagonal coupling. This can similarly be visualised as an obloid with three principle components. Monoclinic and triclinic systems contain three independent diagonal components with one and three non-zero components in the off-diagonals, respectively. In this case, the three-dimensional representation is an obloid with the principle components oriented away from the Cartesian directions. This is presented in Fig.~\ref{fig:dielectric_tensor_form} in tensor and three-dimensional form.

\subsection{Materials Project dielectric dataset}

The dataset used in this work is obtained from the Materials Project (MP).\cite{jain_commentary_2013}
The Materials Project is an open-access database containing c.a.~154,000 material at the time of writing. 
The high-frequency dielectric tensors on the MP are calculated with density functional perturbation theory\cite{lee_high-throughput_2018, ong_materials_2015, perdew_generalized_1996}.
During the dataset cleaning process, we first removed any repeating entries and dielectric tensors with any diagonal elements less than 1.
We further remove any entries with an average polycrystalline dielectric constant greater than 15 as they are unrepresentative of the bulk of the dataset.
In this regime, the training points are very sparse with some outliers possessing extremely high dielectric constants up to 80 that can skew the model training for smaller dielectric constants.
We refer to this dataset as the ``MP-dielectric dataset'' in the rest of this work.
The final dataset is relatively small, containing 6,706 non-repeating entries.
The MP-dielectric dataset is split randomly into training, validation, and test sets, with ratios of 80:10:10.
The training set is used to optimise the model parameters, the validation set is used to determine when to stop the training and for hyperparameter tuning, while model performance is evaluated using the test set.

\begin{figure}[t]
  \centering
  \includegraphics[width=\textwidth]{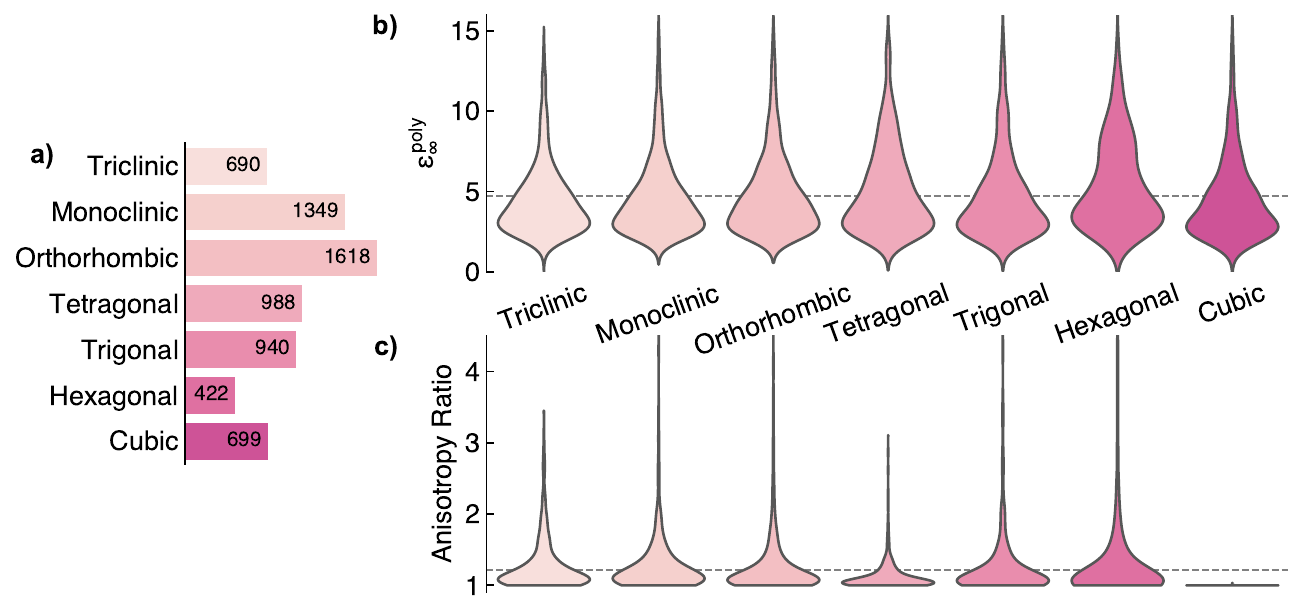}
\caption{Crystal system distribution of Materials Project dielectric dataset. a) Number of materials per crystal system. b,c) Violin plot of polycrystalline dielectric constant, $\varepsilon_\infty^\mathrm{poly}$ (b) and anisotropy ratio (c) across crystal crystal systems. A large area of the violin indicates a greater number of materials with that value. The dashed lines in panels b) and c) indicate the average value across the entire dataset.}
    \label{fig:dataset-distribution}
\end{figure}

After pre-processing, the polycrystalline dielectric constant ($\varepsilon_\infty^\mathrm{poly}$) is obtained by taking the average of the three eigenvalues of the dielectric tensor.
The breakdown across crystal system is presented in Fig.~\ref{fig:dataset-distribution}a.
The dataset is somewhat unbalanced, with over 3 times as many orthorhombic crystals present as hexagonal systems.
The distribution of $\varepsilon_\infty^\mathrm{poly}$ across crystal system is highlighted in Fig.~\ref{fig:dataset-distribution}b, with each class displaying a roughly similar distribution of dielectric values.
For each symmetry type, the histogram depicts a skewed distribution, with a peak in frequency around 3.5 and long tails reaching up to an  $\varepsilon_\infty^\mathrm{poly}$ of 15.
The average polycrystalline dielectric constant across the entire dataset is 4.74.
To evaluate how well our model captures anisotropy, we propose the anisotropy ratio metric, $a_\mathrm{r}$, obtained as the ratio of the largest and smallest dielectric tensor eigenvalues. 
A larger $a_\mathrm{r}$ indicates a more anisotropic system, with an $a_\mathrm{r}$ of 1 indicating a completely isotropic material.
Fig \ref{fig:dataset-distribution}c displays the distribution of $a_\mathrm{r}$ in the training dataset sorted by crystal system.
In all crystal systems, the distribution of the $a_\mathrm{r}$ is heavily skewed towards 1 with long tails reaching up to around 4.
An exception is for cubic symmetry which, by definition, is isotropic and therefore has a value of exactly 1.
There are several outliers for each crystal system, with an $a_\mathrm{r}$ of up to 12.
Most of these materials are two-dimensional (2D) layered materials.
The most anisotropic system of all is hexagonal, with an average $a_\mathrm{r}$ of 1.28, followed closely by triclinic at 1.27.
The average anistropy ratio across all systems is 1.22.

\subsection{Model details and architecture}

AnisoNet is an equivariant message-passing graph neural network\cite{zhou_graph_2020} built on the \textsc{e3nn}\cite{geiger_e3nn_2022} and \textsc{PyTorch}\cite{paszke2017automatic} libraries. AnisoNet takes a periodic crystal graph as input and performs several message-passing steps, before outputting the irreducible representation of the dielectric tensor, which is trivially converted to its Cartesian form via Eqn.~\ref{eq:decomposition}. The basic framework for our approach has been widely used in materials science to predict properties such as elastic tensors, energies, and charge densities\cite{jorgensen_equivariant_2022, wen_equivariant_2024, batatia_mace_2023}. Atoms are represented by nodes, with node features $\mathbf{F}$ and attributes $\mathbf{A}$. Node features are continuously updated during the message-passing steps, whereas attributes remain fixed throughout. Initial node features are generated from a 118-long vector (where the $Z$-th component is the atomic mass of the $Z$-th element) which is passed through an embedding layer. The node attributes are generated using a similar approach but using a one-hot encoding scheme (employing $1$s rather than the atomic mass). The same embedding layer is used for both node features and attributes, as previously employed by \citet{chen_phonon_2019}. We found this approach to yield better performance than using either separate embeddings or the same vectors for both features and attributes. A cutoff of \SI{5}{\angstrom} is employed for graph edge construction, taking into account periodic boundary conditions. An edge vector, $\mathbf{r}_{ij}$, between two nodes $i$ and $j$ is represented by its length $||\textbf{r}_{ij}||$ and unit vector $\mathbf{\hat{r}}_{ij}$. The edge length is projected onto 15 equally spaced Gaussians and embedded using a linear layer, while the edge unit vector is projected onto spherical harmonics up to the maximum rotational order defined by the model hyperparameter $l_\mathrm{max}$.
Through hyperparameter tuning, we identified an optimal $l_\mathrm{max}$ of 3 for AnisoNet.

Our implementation broadly follows the design of Tensor Field Networks\cite{thomas_tensor_2018} and NequIP\cite{batzner_e3-equivariant_2022}, along with the work of \citet{chen_phonon_2019}
In contrast to other graph neural networks for molecules and materials that employ scalar features\cite{chen_graph_2019, chen_universal_2022, sanyal_mt-cgcnn_2018, schutt_schnet_2018}, in AnisoNet the atom features are a series of scalars, vectors, and higher-rank tensors.
Together, the features can be seen as a geometric object comprised of a direct sum of irreducible representations of the SO(3) group\cite{backus_geometrical_1970}.
As previously discussed, geometric features have the benefit that they act as an inductive bias that reduce the amount of training data by avoiding data augmentation and improve model performance\cite{smidt_finding_2021}.
Furthermore, as in the case of this work, they are naturally suited for constructing physical tensors such as the dielectric constant.

\begin{figure}[t]
  \centering
    \centering
    \includegraphics[width=0.65\linewidth]{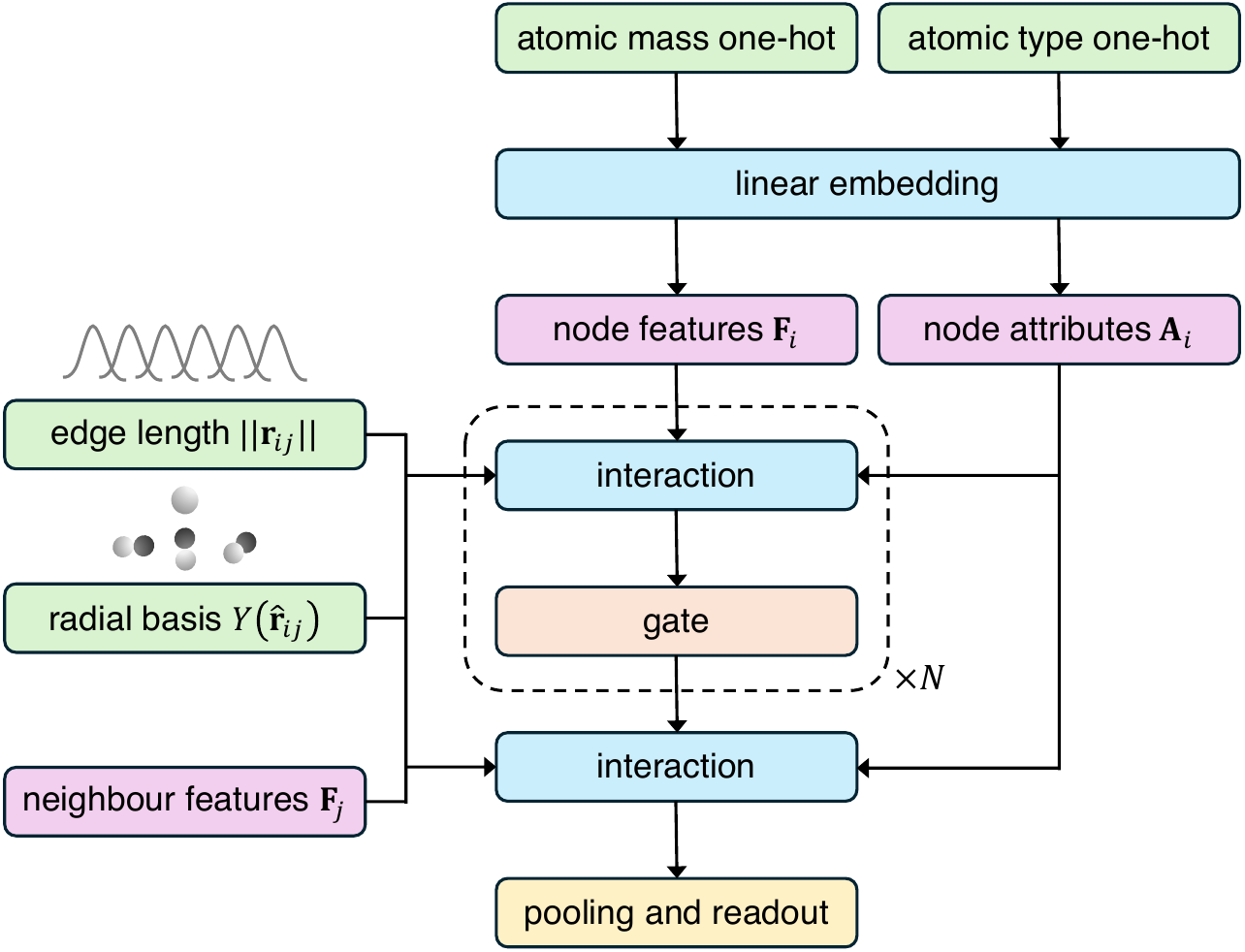}
    \caption{High-level overview of the architecture of AnisoNet. Blue boxes indicate blocks with trainable weights, green boxes indicate model inputs, and purple boxes are model features.}
    \label{fig:model_architecture}
\end{figure}

With each message passing step $t$, the node features are  updated using an equivariant interaction function. The message-passing operation $\mathcal{L}$ in AnisoNet is defined as
\begin{equation}
\mathcal{L}^{L_3}_{\text{acm}_3} \left(\mathbf{r}_{ij}, F_{\text{jcm}_i}\right) \coloneq \sum_{m_1, m_2} C^{L_3, m_3}_{L_1, m_1; L_2, m_2} \sum_{j \in \mathcal{N}(i)} R_c^{l_1, l_2}(||\textbf{r}_{ij}||) Y_{l_1}^{m_1} (\mathbf{\hat{r}}_{ij}) F_{\text{jcm}_1}^{l_1},
\end{equation}
where $C$ is the Clebsch-Gordon coefficient which determines if two spherical harmonics have the correct symmetry to interact, $j\in\mathcal{N}(i)$ represents the set of neighbours of node $i$, $R$ denotes a learnable multilayer perceptron (MLP) acting on the radial basis projection of the edge length, $Y$ is a spherical harmonic basis function, $c$ is the channel index, and $m$ is the index for a degree-$L$ spherical harmonic taking the values $m=-l, \dots, l$. The coefficient $C$ is finite if two features possess the correct symmetry to interact and zero otherwise.
The message passing is a generalised tensor product that interacts geometric features in a rotational equivariant manner, defined by the outer product of two vectors, followed by harmonic (or Wigner) decomposition.\cite{backus_geometrical_1970}.

Following \citet{chen_phonon_2019}, in all but the last message passing step, the node features are passed through a gated non-linearity. A ReLU non-linearity is applied to scalar features, while each tensor feature is associated with an additional scalar feature which is itself passed through a ReLU and used to scale the tensor.  Following the final message passing step, the node features are pooled to obtain the final dielectric irreducible representation output as
\begin{equation}
    \bm{\varepsilon} = \frac{1}{N}\sum^{N}_{i}\mathbf{F}_i,
\end{equation}
where $N$ is the total number of atoms in the structure. The output of the model consists of one scalar and one symmetric traceless tensor, denoted as 0e+2e in \textsc{e3nn} notation.
The final message passing step will only use $l = 0$ and $l= 2$ features to construct the output, while preceding message passing steps will make use of all features up to $l_\mathrm{max}$.

\subsection{Scalar dielectric model architecture}

As a comparison to AnisoNet, we trained an additional model with the output irreducible representation set to 6$\times$0e.
In this case, the model predicts the 6 independent terms of the dielectric tensor separately.
We refer to this model as the ``scalar model'', as it aims to demonstrate the effect of equivariance on model performance. 
Unlike AnisoNet, the scalar model does not have the inductive bias constraining the symmetry of the output.
For example, by construction, the equivariant model cannot produce a non-cubic dielectric tensor given a cubic input structure. 
In contrast, the scalar model has no such restrictions and, as we shall demonstrate, often predicts dielectric tensors with multiple independent components even for cubic systems.
We note that equivariance is only broken in the final read-out stage, with equivariance still retained in preceding message-passing steps.
As part of model testing, we train a scalar model with $l_\mathrm{max} = 0$.
This can be seen as a fully invariant (rather than equivariant) model since equivariance is not enforced at any stage of the network.
The optimal $l_\mathrm{max}$ for the scalar model was identified as 2 through hyperparameter tuning.

\subsection{Model training and hyperparameter optimisation}

Training and hyperparameter tuning of AnisoNet were performed on nodes with either 4 Nvidia GTX 1080Ti or 2 Nvidia TitanX.
We used the AdamW optimizer\cite{loshchilov_decoupled_2019} to minimise the mean-squared error (MSE) between the predicted and target dielectric irreducible representation.
Notably, neither the dielectric tensor eigenvalues nor anisotropy ratio were involved in the loss computation during training.
Instead, the polycrystalline (spherically averaged) dielectric tensor and $a_\mathrm{r}$ were computed at the end of training. 
An exponential decaying learning rate scheduler with an exponential factor of 0.98 was employed.
This enables the model to reach convergence and prevent overfitting and oscillations of validation loss.
Tensor products and irreducible representation-related computations were performed using the \textsc{e3nn} library\cite{geiger_e3nn_2022}.
To obtain the optimal parameters for AnisoNet, we first performed Bayesian optimisation with the \textsc{optuna} library\cite{akiba_optuna_2019} before a grid search was applied around the optimal parameters identified.
The range of model parameters considered by \textsc{optuna} and the grid search are presented in Table \ref{tab:hyperparameter-tuning}.

\begin{table}[H]
    \centering
    \begin{tabular}{l l l}
        \toprule
        Hyperparameter & BO & Grid Search \\ \midrule
        Learning rate & 0.0001 - 0.5 & 0.002, 0.003, \textbf{0.004}, 0.005 \\
        Embedding length & 8 - 96 & 16, 32, \textbf{48}, 64 \\
        Number of gate layers & 1, 2, 3, 4 & 1, \textbf{2}, 3 \\
        Irrep multiplicity & 16 - 64 & 16, 32, \textbf{48} \\
        Maximum Rotational Order $l_\mathrm{max}$ & 1, 2, 3, 4 & 1, 2, \textbf{3} \\ \bottomrule
    \end{tabular}
    \caption{The ranges considered for hyperparameter tuning during Bayesian optimisation with the \textsc{Optuna}\cite{akiba_optuna_2019} package and grid search.}
    \label{tab:hyperparameter-tuning}
\end{table}

\subsection{High-throughput density functional theory calculations}

\textit{Ab initio} calculations were performed using the Vienna \textit{ab initio} Simulation Package (VASP),\cite{kresse_initio_1993a,kresse_efficiency_1996} a planewave density functional theory package.
Calculations were orchestrated in a high-throughput mode using the \textsc{atomate2} software,\cite{Ganose_atomate2_2024} with workflows written and executed using the \textsc{jobflow}\cite{Rosen_Jobflow_Computational_Workflows_2024} and \textsc{jobflow-remote}\cite{jobflow_remote} libraries.
The calculations follow the Materials Project\cite{jain_commentary_2013} input settings to ensure consistency with the original dataset using the \verb|MPStaticSet| and \verb|MPNonSCFSet| classes in the pymatgen library.\cite{ong_python_2013}
This includes the PBE exchange-correlation functional\cite{perdew_generalized_1996} and a planewave energy cutoff of \SI{520}{\eV}. To achieve sufficient accuracy, we increase the k-point mesh sampling to \SI{200}{\per\cubic\angstrom} and use a tighter energy convergence criterion of \SI{1e-5}{\eV}.
The high-frequency dielectric constant was obtained from density functional perturbation theory (DFPT).\cite{gajdos_linear_2006}
Optical absorption was calculated through the frequency-dependent microscopic polarisability matrix as implemented in VASP.\cite{gajdos_linear_2006}

\section{Results and Discussion}

\subsection{Polycrystalline dielectric constant}

\begin{figure}[t]
\centering
\includegraphics[width=\textwidth]{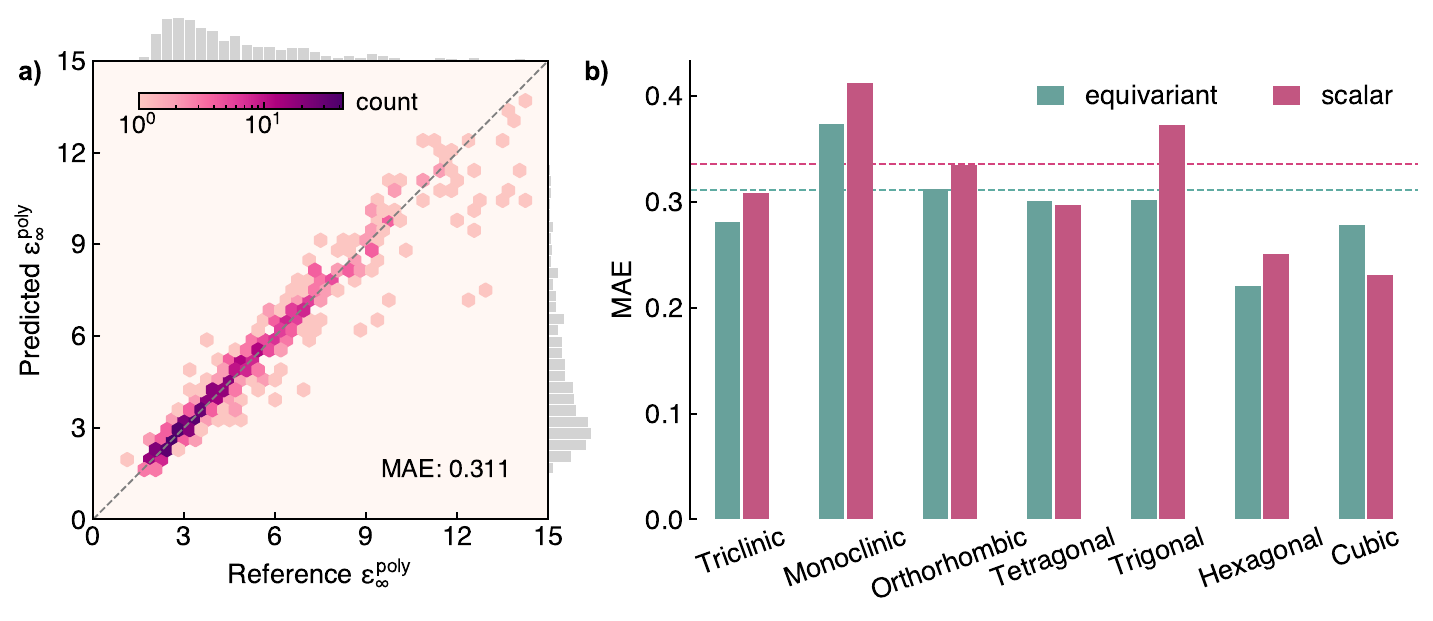}
\caption{Performance of AnisoNet at predicting the polycyrstalline dielectric constant, $\epsilon_\infty^\mathrm{poly}$, on the Materials Project dielectric test set. a) Heatmap of reference vs predicted dielectric constants where a darker colour indicates more materials. b) Comparison of equivariant (0e+2e) and scalar (6x0e) models across crystal systems. The dashed lines represent the average MAE for both models.}
\label{fig:scalar_mae}
\end{figure}

AnisoNet predicts the full 3$\times$3 dielectric tensor.
To test the performance of the model, we first evaluate it on the polycrystalline dielectric constant, $\epsilon_\infty^\mathrm{poly}$.
As detailed in the Methodology, this is obtained by taking the average of the eigenvalues of the dielectric tensor.
We find the predictions of the model agree well with the values calculated using DFPT, with a mean absolute error of 0.311 (Fig.~\ref{fig:scalar_mae}a).
The average dielectric constant across the test dataset is 4.6, giving a percentage error of \SI{6.6}{\percent} (the performance on the training and validation sets are presented in Figure S1 of the Supplementary Information).
For most materials discovery tasks, this error is relatively small and comparable to the variation seen across different exchange-correlation functionals.\cite{petousis2016benchmarking}
The errors are relatively uniformly distributed across the entire range from 1 to 15, albeit with increased deviation from DFT for systems with higher dielectric constants.
This is expected due to the unbalanced training dataset, with less than \SI{5}{\percent} of samples possessing dielectric constants greater than 10.
The MatBench\cite{dunn_benchmarking_2020} dataset and leaderboard have emerged as standard benchmarks for machine learning in materials science.
The performance of AnisoNet is comparable to the best-performing graph neural network, CoGN\cite{ruff_connectivity_2024}, on the MatBench dielectric dataset, with an MAE of 0.309.
However, we note the MAE values are not strictly comparable since we do not use the same training dataset and MatBench uses a different nested cross-validation scheme to avoid bias due to the choice of test set.
A further critical distinction of AnisoNet is its ability to predict the complete dielectric tensor, as opposed to just a scalar value, as is common among all properties listed on the MatBench leaderboard.
Regardless, we find our model achieves suitable accuracy for materials discovery.

To understand the impact of equivariance on model performance, we trained an analogous model with the output irreducible representation set to 6$\times$0e.
In this case, the model predicts six scalars, corresponding to the six independent components of the dielectric tensor.
We compare the performance of the scalar and equivariant (0e+2e) models in Fig.~\ref{fig:scalar_mae}b as a function of crystal system.
The scalar model performs marginally worse than AnisoNet, with an MAE of 0.336.
The errors are relatively consistent across crystal symmetry, with the scalar model having slightly larger errors for all  systems except for cubic space groups.
The similar accuracy of both models indicates that the main contribution to the good performance stems from the invariant features, such as composition and edge length. 

\subsection{Anisotropic dielectric tensor}

While the polycrystalline dielectric constant is sufficient to describe the response of isotropic materials, most crystals are inherently anisotropic and thus the dielectric response will be dependent on the direction it is measured.
Anisotropy results from a crystal's structure including the Bravais lattice type and arrangement of the atoms.
As Neumann's principle states,\cite{neumann1885vorlesungen} the symmetry operations of any physical property must include the symmetry operations of the point group of the crystal exhibiting that property.
To investigate the ability of our model to capture anisotropic properties, we evaluate AnisoNet on the anisotropy ratio, $\alpha_\mathrm{r}$ metic, defined as the quotient of the smallest and largest eigenvalues of the dielectric tensor.
By definition, the minimum possible $a_\mathrm{r}$ is 1 for isotropic systems, with most materials in our dataset possessing values between 1--5.
The distribution of $a_\mathrm{r}$ is heavily skewed towards 1, with over \SI{90}{\percent} of materials exhibiting an $a_\mathrm{r}$ of less than 1.56. 

\begin{figure}[t]
\centering
\includegraphics[width=\textwidth]{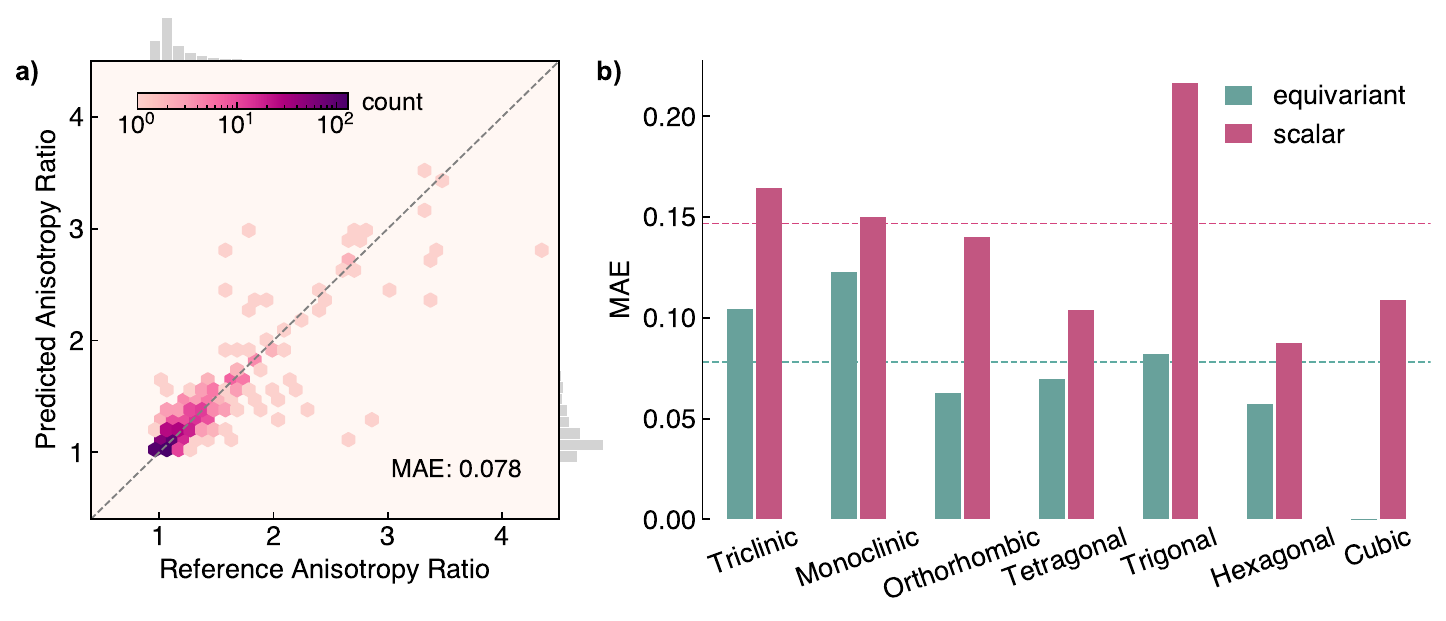}
\caption{Performance of AnisoNet at predicting the anisotropy ratio, $a_\mathrm{r}$, on the Materials Project dielectric test set. a) Heatmap of reference vs predicted values where a darker colour indicates more materials. b) Comparison of equivariant (0e+2e) and scalar (6x0e) models across crystal systems. The dashed lines represent the average MAE for both models.}
\label{fig:anisotropy_ratio}
\end{figure}

AnisoNet achieves an MAE of 0.078 for the 671 materials in the test set, representing a percentage error of \SI{5.96}{\percent} (the performance on the training and validation sets are presented in Figure S2 of the Supplmentary Information).
This indicates relatively good performance, especially considering that $a_\mathrm{r}$ was not directly employed in the loss function during model training.
Despite this, $a_\mathrm{r}$ can be derived from the dielectric irreducible representation on which the model was trained, thereby enabling the model to learn this property indirectly. 
In Fig.~\ref{fig:anisotropy_ratio}a, we visualise the performance of our model against the DFPT reference values.
The predictions show significantly larger deviations above an $a_\mathrm{r}$ of 2.
Through analysis of these errors, we find the large discrepancies are largely caused by the difficulty of our model in predicting large dielectric tensors due to the imbalanced nature of the training set.
An alternative approach to constructing a less sensitive anisotropy metric could be to use the difference between the largest and smallest eigenvalues.
However, for many technological applications such as emerging dark matter detectors,\cite{griffin_multichannel_2020} the relevant property of interest is the differential anisotropy along the crystal axes.
As such, we believe our metric is more appropriate for materials screening.

To elucidate the impact of equivariance on predictions of anisotropy, we compare against the scalar 6x0e model.
In contrast to the polycrystalline dielectric constant, for anisotropy ratio, the scalar model performs considerably worse across all crystal systems (Fig.~\ref{fig:anisotropy_ratio}b).
The MAE of the scalar model is 0.147, almost twice that of equivariant AnisoNet.
Our results highlight the difficulty of learning tensorial properties compared to invariant scalars for traditional machine learning approaches.
While for AnisoNet, the errors are typically smaller for more symmetric systems due to the inductive bias provided by symmetry, the scalar model displays no such trend, with the largest MAE of 0.217 seen for tricilinic space groups.
Notably, the MAE on cubic systems is 0.108 for the scalar model but exactly zero for AnisoNet due to symmetry constraints. 
We stress that the scalar model is still internally equivariant up to the final readout stage and is able to capture angular features due to the use of spherical harmonic descriptors ($l_\mathrm{max}$ of 2) for the edge feature vectors. 

\subsection{Impact of angular features on model performance}

To further understand the impact of the angular resolution on model performance, we train a series of models with increasing $l_\mathrm{max}$ (from 0 to 4) for both the equivariant and scalar models.
The full results are presented in Table \ref{tab:AR_lmax}.
The equivariant model outperforms the scalar model across all $l_\mathrm{max}$ values.
As previously discussed, the difference is marginal for $\varepsilon^\mathrm{poly}_\infty$ (average improvement of \SI{8}{\percent}) but more noticeable for $a_\mathrm{r}$ (average improvement of \SI{58}{\percent}).
This reinforces our conclusion that invariant features are the dominant factor driving the predictions of the polycrystalline dielectric constant.

Both models show a similar trend with increasing $l_\mathrm{max}$.
The mean average errors initially fall due to the greater ability of the models to capture finer angular features.
However, at larger $l_\mathrm{max}$, the model errors begin to increase due to the greater number of model parameters that lead to underfitting with the relatively small dataset size.
This trend is found for both equivariant and scalar models and across both the polycrystalline dielectric constant and anisotropy ratio tasks.
Notably, we find that an $l_\mathrm{max}$ greater or equal to 2 is essential for the equivariant model to achieve good performance for anisotropy ratio.
This is to be expected since the irreducible representation of the Cartesian dielectric tensor (0e+2e) contains an $l=2$ component. 

\begin{table}[H]
    \centering
    \begin{tabular}{ccccc}
        \toprule
        & \multicolumn{2}{c}{$\varepsilon_\infty^\mathrm{poly}$ MAE} & \multicolumn{2}{c}{$a_\mathrm{r}$ MAE} \\ \cmidrule(lr){2-3}\cmidrule(lr){4-5}
        $l_\mathrm{max}$ & Equivariant & Scalar & Equivariant & Scalar \\\midrule
        0 & -- & 0.439 & -- & 0.184 \\
        1 & 0.327 & 0.390 & 0.130 & 0.168 \\
        2 & 0.335 & 0.336 & 0.095 & 0.147 \\
        3 & \textbf{0.311} & 0.345 & \textbf{0.078} & 0.149\\
        4 & 0.340 & -- & 0.087 & -- \\ \bottomrule
    \end{tabular}
    \caption{Impact of maximum rotational order $l_\mathrm{max}$ on model performance. Mean average error (MAE) of equivariant (0e+2e) and scalar (6$\times$0e) models on the polycrystalline dielectric constant ($\varepsilon_\infty^\mathrm{poly}$) and anisotropy ratio ($a_\mathrm{r}$). The best model performance is highlighted in bold.}
    \label{tab:AR_lmax}
\end{table}

\subsection{Discovery of highly anisotropic dielectric crystals}

\begin{figure}[t]
  \centering
  \includegraphics[width=0.9\textwidth]{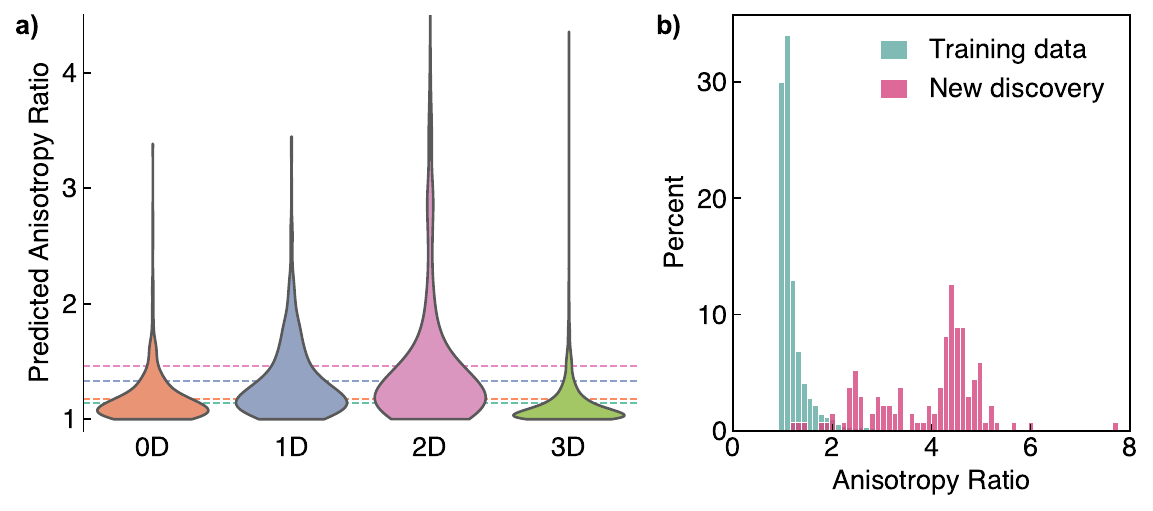}
  \caption{Novel ansiotropic materials discovered by high-throughput density functional perturbation theory calculations. a) Distribution of anisotropy ratio with respect to the structural dimensionality calculated using robocrystallographer.\cite{Ganose_Jain_2019} The dashed lines indicate the average for each dimensionality class. b) Histogram of ansiotropy ratio for the training dataset and new materials calculated.}
  \label{fig:mp-histogram}
\end{figure}

Crystalline materials with anisotropic dielectric properties are of interest as optical fibre sensors, linear optical devices, and in advanced optical communication (due to their birefringence),\cite{tudi2022potential} along with a host of exotic applications such as dark matter detectors\cite{griffin_multichannel_2020, coskuner_directional_2021} and beyond Moore's law computing.\cite{liu2021coherent}
We applied AnisoNet to search for materials with large $a_\mathrm{r}$.
To begin, we filtered the Materials Project\cite{jain_commentary_2013} for structures with the following criteria: i) An energy above the hull of less than \SI{50}{\meV/atom} to select thermodynamically stable candidates; ii) Exclusion of materials purely consisting of noble gases or hydrogen due to the lack of these elements in the training set (in our testing these system led to unphysical predictions with eigenvalues less than 1); iii) Structures with less than 40 sites in a unit cell, to enable validation of candidates with DFPT calculations; iv) A band gap greater than \SI{0.5}{\eV}, since the dielectric response is extremely sensitive in narrow gap semiconductors; and v) The exclusion of any materials that already exist in the MP-dielectric dataset.
After filtering, we obtained a set of 18,835 structures for which we obtained the dielectric tensor using AnisoNet.
An analysis of the predicted polycrystalline dielectric constants along with the impact of chemical composition are provided in Section S2 of the Supplementary Information.
To understand the impact of structural connectivity on the dielectric anisotropy, we calculated the dimensionality of each structure using the robocrystallographer\cite{Ganose_Jain_2019} and CrystalNN packages.\cite{pan2021benchmarking}
Figure \ref{fig:mp-histogram}a illustrates the distribution of $a_\mathrm{r}$ with respect to dimensionality.
Structures with three-dimensional (3D) connectivity exhibit the lowest anisotropy, with an average $a_\mathrm{r}$ of 1.1. 
Notably, over \SI{80}{\percent} of the materials in our dataset are 3D, making the identification of any anisotropic 3D materials particularly valuable.
In contrast, 2D and 1D materials display the greatest degree of anisotropy with average anisotropy ratios of 1.5 and 1.3, respectively, as expected due to the presence of layers or ribbons in the structures.

We select all materials with an $a_\mathrm{r}$ greater than 2.5 for validation using high-throughput DFPT calculations using the \textsc{atomate2} package.\cite{Ganose_atomate2_2024}
This amounts to 137 structures, mostly comprising 2D candidates (107 materials) with fewer 3D (23 materials), 1D (5 materials), and 0D (2 materials) structures.
Almost all of the 2D structures are transition metal dichalcogenides containing tungsten or molybdenum.
Furthermore, upon closer inspection, many of the 3D materials could also be classed as pseudo-two-dimensional, since they contain layered components with intercalated spectator ions such as lithium or sodium.
Fig.~\ref{fig:mp-histogram}b presents a histogram of $a_\mathrm{r}$ for the calculated materials in comparison to the training dataset.
Over \SI{95}{\percent} of the new discoveries have a calculated $a_\mathrm{r}$ of over 2, with an average value of 3.9 in contrast to an average of 1.2 for the structures in the training data.
This highlights the effectiveness of AnisoNet in searching for highly anisotropic dielectric properties.
The dielectric tensors for all identified crystals are provided in Table S1 of the Supplementary Information.

\begin{figure}[t]
  \centering
  \includegraphics[width=\textwidth]{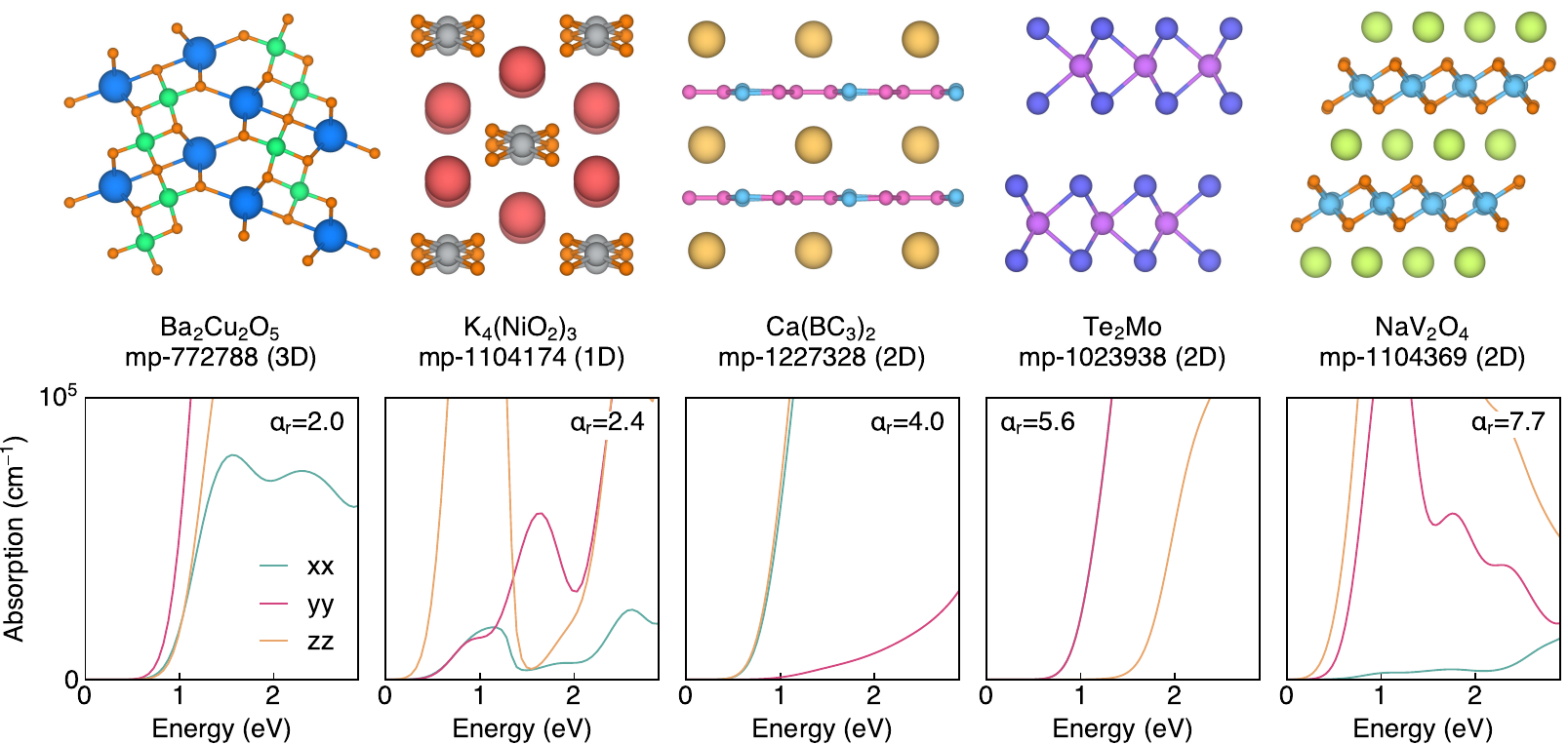}
\caption{Crystal structures and optical absorption spectra for a selection of highly anisotropy materials identified by our screening.}
    \label{fig:optical-absorption}
\end{figure}

We further select 5 materials for which we obtain the full frequency-dependent dielectric constant.
These include \ce{NaV2O4} which has the highest $a_\mathrm{r}$ of 7.74 in our dataset.
The crystal structure, presented in Fig.~\ref{fig:optical-absorption}, contains edge-sharing \ce{VO6} octahedral layers separated by interstital sodium ions.
The optical absorption spectrum exhibits giant anisotropy, with the absorption in the directions parallel to the \ce{VO6} sheets quickly reaching \SI{e5}{\per\cm}, while the perpendicular direction has an absorption close to zero.
\ce{K4N(NiO2)3} is a 1D material containing edge sharing \ce{NiO4} square planar polyhedra separated by potassium counter ions.
The optical absorption spectrum shows a complex directional dependence that varies with excitation energy, with strong anisotropy observed at energies of 0.9 and \SI{2.5}{\eV}.
Finally, \ce{Ba2Cu2O5} crystallises in a complex structure containing edge-sharing square planar Cu$_2$O$_6$ dimers and edge-sharing BaO$_6$ octahedra, with an $a_\mathrm{r}$ of 1.97 (the largest for a 3D connected material in our dataset).
At energies close to the band gap, the absorption is relatively isotropic but gains significant anistropy for shorter wavelengths.
These results highlight the potential of discovering materials with high optical anisotropy using the high-frequency dielectric constant as a proxy.

\section{Conclusion}

In this work, we developed AnisoNet, an equivariant graph neural network to predict the full dielectric tensors of crystalline materials.
The predictions of AnisoNet will always be consistent with the input structure symmetry, therefore eliminating unphysical tensors such as anisotropy for cubic materials.
To quantify the degree of anisotropy we introduced a metric termed the anisotropy ratio.
We demonstrated that AnisoNet has comparable performance to state-of-the-art models whilst also being able to capture the full directional response.
AnisoNet was applied to screen the Materials Project dataset for highly anisotropic dielectric crystals, through which we identified 137 materials with large anisotropy ratios, some of which possessed three-dimensional connectivity.
Our results highlight the efficiency of equivariant graph neural networks for the discovery of novel materials with tailored tensorial properties.

An obvious limitation of AnisoNet is the use of a cutoff radius of \SI{5}{\angstrom} when constructing the graph.
Lower-dimensional structures containing weak van der Waals with large separations between structural components are often of interest for their anisotropic properties.
Indeed, our work indicated that \SI{80}{\percent} of materials with large predicted $\alpha_\mathrm{r}$ are 2D or quasi-2D structures.
As AnisoNet will not be capable of distinguishing large gaps or voids if the separation is larger than the cutoff, alternative graph construction methods may yield better results, such as graph rewiring as developed by \citet{di_giovanni_over-squashing_2023}
However, such an approach must be adapted to ensure that the equivariant properties of the network are retained.

\section{Data access statement}

Model weights, datasets and scripts used to generate all results and figures shown in this work can be found at \url{https://github.com/virtualatoms/AnisoNet}.

\section{Conflicts of Interest}

There are no conflicts to declare.

\section{Acknowledgements}

We thank Jason Munro for help with obtaining the dielectric tensor dataset from the Materials Project. A.M.G.~was supported by EPSRC Fellowship EP/T033231/1. We are grateful to the UK Materials and Molecular Modelling Hub for computational resources, which are partially funded by EPSRC (EP/T022213/1, EP/W032260/1 and EP/P020194/1). This project made use of time on the Tier 2 HPC facility JADE, funded by EPSRC (EP/P020275/1).

\bibliography{refs}

\providecommand{\latin}[1]{#1}
\makeatletter
\providecommand{\doi}
  {\begingroup\let\do\@makeother\dospecials
  \catcode`\{=1 \catcode`\}=2 \doi@aux}
\providecommand{\doi@aux}[1]{\endgroup\texttt{#1}}
\makeatother
\providecommand*\mcitethebibliography{\thebibliography}
\csname @ifundefined\endcsname{endmcitethebibliography}  {\let\endmcitethebibliography\endthebibliography}{}
\begin{mcitethebibliography}{75}
\providecommand*\natexlab[1]{#1}
\providecommand*\mciteSetBstSublistMode[1]{}
\providecommand*\mciteSetBstMaxWidthForm[2]{}
\providecommand*\mciteBstWouldAddEndPuncttrue
  {\def\EndOfBibitem{\unskip.}}
\providecommand*\mciteBstWouldAddEndPunctfalse
  {\let\EndOfBibitem\relax}
\providecommand*\mciteSetBstMidEndSepPunct[3]{}
\providecommand*\mciteSetBstSublistLabelBeginEnd[3]{}
\providecommand*\EndOfBibitem{}
\mciteSetBstSublistMode{f}
\mciteSetBstMaxWidthForm{subitem}{(\alph{mcitesubitemcount})}
\mciteSetBstSublistLabelBeginEnd
  {\mcitemaxwidthsubitemform\space}
  {\relax}
  {\relax}

\bibitem[Ohno(2010)]{ohno_window_2010}
Ohno,~H. A window on the future of spintronics. \emph{Nature Materials} \textbf{2010}, \emph{9}, 952--954\relax
\mciteBstWouldAddEndPuncttrue
\mciteSetBstMidEndSepPunct{\mcitedefaultmidpunct}
{\mcitedefaultendpunct}{\mcitedefaultseppunct}\relax
\EndOfBibitem
\bibitem[Oberg \latin{et~al.}(2014)Oberg, Calvo, Delgado, Moro-Lagares, Serrate, Jacob, Fernández-Rossier, and Hirjibehedin]{oberg_control_2014}
Oberg,~J.~C.; Calvo,~M.~R.; Delgado,~F.; Moro-Lagares,~M.; Serrate,~D.; Jacob,~D.; Fernández-Rossier,~J.; Hirjibehedin,~C.~F. Control of single-spin magnetic anisotropy by exchange coupling. \emph{Nature Nanotechnology} \textbf{2014}, \emph{9}, 64--68\relax
\mciteBstWouldAddEndPuncttrue
\mciteSetBstMidEndSepPunct{\mcitedefaultmidpunct}
{\mcitedefaultendpunct}{\mcitedefaultseppunct}\relax
\EndOfBibitem
\bibitem[Li \latin{et~al.}(2021)Li, Hu, He, Yao, Fu, Wang, Zhao, Felser, and Zhu]{li_demonstration_2021}
Li,~A.; Hu,~C.; He,~B.; Yao,~M.; Fu,~C.; Wang,~Y.; Zhao,~X.; Felser,~C.; Zhu,~T. Demonstration of valley anisotropy utilized to enhance the thermoelectric power factor. \emph{Nature Communications} \textbf{2021}, \emph{12}, 5408\relax
\mciteBstWouldAddEndPuncttrue
\mciteSetBstMidEndSepPunct{\mcitedefaultmidpunct}
{\mcitedefaultendpunct}{\mcitedefaultseppunct}\relax
\EndOfBibitem
\bibitem[Li \latin{et~al.}(2023)Li, Huang, Zhou, Xu, Qin, Chen, Sun, Liu, Sui, Qiu, Lu, Gou, Xi, Ideue, Tang, Iwasa, and Yuan]{li_anisotropic_2023}
Li,~Z. \latin{et~al.}  An anisotropic van der {Waals} dielectric for symmetry engineering in functionalized heterointerfaces. \emph{Nature Communications} \textbf{2023}, \emph{14}, 5568\relax
\mciteBstWouldAddEndPuncttrue
\mciteSetBstMidEndSepPunct{\mcitedefaultmidpunct}
{\mcitedefaultendpunct}{\mcitedefaultseppunct}\relax
\EndOfBibitem
\bibitem[Pan and Randall(2010)Pan, and Randall]{pan_brief_2010}
Pan,~M.-J.; Randall,~C.~A. A brief introduction to ceramic capacitors. \emph{IEEE Electrical Insulation Magazine} \textbf{2010}, \emph{26}, 44--50\relax
\mciteBstWouldAddEndPuncttrue
\mciteSetBstMidEndSepPunct{\mcitedefaultmidpunct}
{\mcitedefaultendpunct}{\mcitedefaultseppunct}\relax
\EndOfBibitem
\bibitem[Sun \latin{et~al.}(2022)Sun, Luo, and Li]{sun_review_2022}
Sun,~J.; Luo,~B.; Li,~H. A {Review} on the {Conventional} {Capacitors}, {Supercapacitors}, and {Emerging} {Hybrid} {Ion} {Capacitors}: {Past}, {Present}, and {Future}. \emph{Advanced Energy and Sustainability Research} \textbf{2022}, \emph{3}, 2100191\relax
\mciteBstWouldAddEndPuncttrue
\mciteSetBstMidEndSepPunct{\mcitedefaultmidpunct}
{\mcitedefaultendpunct}{\mcitedefaultseppunct}\relax
\EndOfBibitem
\bibitem[Li \latin{et~al.}(2021)Li, Liu, Ke, Tang, Liu, Huang, Wu, Wu, and Kang]{li_review_2021}
Li,~X.; Liu,~H.; Ke,~C.; Tang,~W.; Liu,~M.; Huang,~F.; Wu,~Y.; Wu,~Z.; Kang,~J. Review of {Anisotropic} {2D} {Materials}: {Controlled} {Growth}, {Optical} {Anisotropy} {Modulation}, and {Photonic} {Applications}. \emph{Laser \& Photonics Reviews} \textbf{2021}, \emph{15}, 2100322\relax
\mciteBstWouldAddEndPuncttrue
\mciteSetBstMidEndSepPunct{\mcitedefaultmidpunct}
{\mcitedefaultendpunct}{\mcitedefaultseppunct}\relax
\EndOfBibitem
\bibitem[Palmer \latin{et~al.}(2014)Palmer, Edwards-Gau, Kariuki, Harris, Dolbnya, and Collins]{palmer_x-ray_2014}
Palmer,~B.~A.; Edwards-Gau,~G.~R.; Kariuki,~B.~M.; Harris,~K. D.~M.; Dolbnya,~I.~P.; Collins,~S.~P. X-ray birefringence imaging. \emph{Science} \textbf{2014}, \emph{344}, 1013--1016\relax
\mciteBstWouldAddEndPuncttrue
\mciteSetBstMidEndSepPunct{\mcitedefaultmidpunct}
{\mcitedefaultendpunct}{\mcitedefaultseppunct}\relax
\EndOfBibitem
\bibitem[Coskuner \latin{et~al.}(2021)Coskuner, Mitridate, Olivares, and Zurek]{coskuner_directional_2021}
Coskuner,~A.; Mitridate,~A.; Olivares,~A.; Zurek,~K.~M. Directional dark matter detection in anisotropic {Dirac} materials. \emph{Physical Review D} \textbf{2021}, \emph{103}, 016006\relax
\mciteBstWouldAddEndPuncttrue
\mciteSetBstMidEndSepPunct{\mcitedefaultmidpunct}
{\mcitedefaultendpunct}{\mcitedefaultseppunct}\relax
\EndOfBibitem
\bibitem[Griffin \latin{et~al.}(2020)Griffin, Inzani, Trickle, Zhang, and Zurek]{griffin_multichannel_2020}
Griffin,~S.~M.; Inzani,~K.; Trickle,~T.; Zhang,~Z.; Zurek,~K.~M. Multichannel direct detection of light dark matter: {Target} comparison. \emph{Physical Review D} \textbf{2020}, \emph{101}, 055004\relax
\mciteBstWouldAddEndPuncttrue
\mciteSetBstMidEndSepPunct{\mcitedefaultmidpunct}
{\mcitedefaultendpunct}{\mcitedefaultseppunct}\relax
\EndOfBibitem
\bibitem[Nurul~Hidayah and Mariatti(2021)Nurul~Hidayah, and Mariatti]{nurul_hidayah_thermoplastic_2021}
Nurul~Hidayah,~I.; Mariatti,~M. \emph{Reference {Module} in {Materials} {Science} and {Materials} {Engineering}}; Elsevier, 2021\relax
\mciteBstWouldAddEndPuncttrue
\mciteSetBstMidEndSepPunct{\mcitedefaultmidpunct}
{\mcitedefaultendpunct}{\mcitedefaultseppunct}\relax
\EndOfBibitem
\bibitem[Spaldin(2012)]{spaldin_beginners_2012}
Spaldin,~N.~A. A beginner's guide to the modern theory of polarization. \emph{Journal of Solid State Chemistry} \textbf{2012}, \emph{195}, 2--10\relax
\mciteBstWouldAddEndPuncttrue
\mciteSetBstMidEndSepPunct{\mcitedefaultmidpunct}
{\mcitedefaultendpunct}{\mcitedefaultseppunct}\relax
\EndOfBibitem
\bibitem[Izgorodina \latin{et~al.}(2009)Izgorodina, Forsyth, and MacFarlane]{izgorodina_components_2009}
Izgorodina,~E.~I.; Forsyth,~M.; MacFarlane,~D.~R. On the components of the dielectric constants of ionic liquids: ionic polarization? \emph{Physical Chemistry Chemical Physics} \textbf{2009}, \emph{11}, 2452--2458\relax
\mciteBstWouldAddEndPuncttrue
\mciteSetBstMidEndSepPunct{\mcitedefaultmidpunct}
{\mcitedefaultendpunct}{\mcitedefaultseppunct}\relax
\EndOfBibitem
\bibitem[Lejaeghere \latin{et~al.}(2016)Lejaeghere, Bihlmayer, Björkman, Blaha, Blügel, Blum, Caliste, Castelli, Clark, Dal~Corso, de~Gironcoli, Deutsch, Dewhurst, Di~Marco, Draxl, Dułak, Eriksson, Flores-Livas, Garrity, Genovese, Giannozzi, Giantomassi, Goedecker, Gonze, Grånäs, Gross, Gulans, Gygi, Hamann, Hasnip, Holzwarth, Iuşan, Jochym, Jollet, Jones, Kresse, Koepernik, Küçükbenli, Kvashnin, Locht, Lubeck, Marsman, Marzari, Nitzsche, Nordström, Ozaki, Paulatto, Pickard, Poelmans, Probert, Refson, Richter, Rignanese, Saha, Scheffler, Schlipf, Schwarz, Sharma, Tavazza, Thunström, Tkatchenko, Torrent, Vanderbilt, van Setten, Van~Speybroeck, Wills, Yates, Zhang, and Cottenier]{lejaeghere_reproducibility_2016}
Lejaeghere,~K. \latin{et~al.}  Reproducibility in density functional theory calculations of solids. \emph{Science} \textbf{2016}, \emph{351}, aad3000\relax
\mciteBstWouldAddEndPuncttrue
\mciteSetBstMidEndSepPunct{\mcitedefaultmidpunct}
{\mcitedefaultendpunct}{\mcitedefaultseppunct}\relax
\EndOfBibitem
\bibitem[Petousis \latin{et~al.}(2017)Petousis, Mrdjenovich, Ballouz, Liu, Winston, Chen, Graf, Schladt, Persson, and Prinz]{petousis_high-throughput_2017}
Petousis,~I.; Mrdjenovich,~D.; Ballouz,~E.; Liu,~M.; Winston,~D.; Chen,~W.; Graf,~T.; Schladt,~T.~D.; Persson,~K.~A.; Prinz,~F.~B. High-throughput screening of inorganic compounds for the discovery of novel dielectric and optical materials. \emph{Scientific Data} \textbf{2017}, \emph{4}, 160134\relax
\mciteBstWouldAddEndPuncttrue
\mciteSetBstMidEndSepPunct{\mcitedefaultmidpunct}
{\mcitedefaultendpunct}{\mcitedefaultseppunct}\relax
\EndOfBibitem
\bibitem[Jain \latin{et~al.}(2013)Jain, Ong, Hautier, Chen, Richards, Dacek, Cholia, Gunter, Skinner, Ceder, and Persson]{jain_commentary_2013}
Jain,~A.; Ong,~S.~P.; Hautier,~G.; Chen,~W.; Richards,~W.~D.; Dacek,~S.; Cholia,~S.; Gunter,~D.; Skinner,~D.; Ceder,~G.; Persson,~K.~A. Commentary: {The} {Materials} {Project}: {A} materials genome approach to accelerating materials innovation. \emph{APL Materials} \textbf{2013}, \emph{1}, 011002\relax
\mciteBstWouldAddEndPuncttrue
\mciteSetBstMidEndSepPunct{\mcitedefaultmidpunct}
{\mcitedefaultendpunct}{\mcitedefaultseppunct}\relax
\EndOfBibitem
\bibitem[Tawfik \latin{et~al.}(2020)Tawfik, Isayev, Spencer, and Winkler]{tawfik_predicting_2020}
Tawfik,~S.~A.; Isayev,~O.; Spencer,~M. J.~S.; Winkler,~D.~A. Predicting {Thermal} {Properties} of {Crystals} {Using} {Machine} {Learning}. \emph{Advanced Theory and Simulations} \textbf{2020}, \emph{3}, 1900208\relax
\mciteBstWouldAddEndPuncttrue
\mciteSetBstMidEndSepPunct{\mcitedefaultmidpunct}
{\mcitedefaultendpunct}{\mcitedefaultseppunct}\relax
\EndOfBibitem
\bibitem[Butler \latin{et~al.}(2018)Butler, Davies, Cartwright, Isayev, and Walsh]{butler_machine_2018}
Butler,~K.~T.; Davies,~D.~W.; Cartwright,~H.; Isayev,~O.; Walsh,~A. Machine learning for molecular and materials science. \emph{Nature} \textbf{2018}, \emph{559}, 547--555\relax
\mciteBstWouldAddEndPuncttrue
\mciteSetBstMidEndSepPunct{\mcitedefaultmidpunct}
{\mcitedefaultendpunct}{\mcitedefaultseppunct}\relax
\EndOfBibitem
\bibitem[Dawid \latin{et~al.}()Dawid, Arnold, Requena, Gresch, Płodzien, Donatella, Nicoli, Stornati, Koch, Büttner, Okuła, Muñoz, Vargas, Carrasquilla, Dunjko, Gabrié, van Nieuwenburg, Vicentini, Wang, Carleo, Greplová, Krems, Marquardt, Tomza, Lewenstein, and Dauphin]{dawid_modern_nodate}
Dawid,~A. \latin{et~al.}  Modern applications of machine learning in quantum sciences. 283\relax
\mciteBstWouldAddEndPuncttrue
\mciteSetBstMidEndSepPunct{\mcitedefaultmidpunct}
{\mcitedefaultendpunct}{\mcitedefaultseppunct}\relax
\EndOfBibitem
\bibitem[Wu \latin{et~al.}(2021)Wu, Pan, Chen, Long, Zhang, and Yu]{wu_comprehensive_2021}
Wu,~Z.; Pan,~S.; Chen,~F.; Long,~G.; Zhang,~C.; Yu,~P.~S. A {Comprehensive} {Survey} on {Graph} {Neural} {Networks}. \emph{IEEE Transactions on Neural Networks and Learning Systems} \textbf{2021}, \emph{32}, 4--24\relax
\mciteBstWouldAddEndPuncttrue
\mciteSetBstMidEndSepPunct{\mcitedefaultmidpunct}
{\mcitedefaultendpunct}{\mcitedefaultseppunct}\relax
\EndOfBibitem
\bibitem[Bronstein \latin{et~al.}(2021)Bronstein, Bruna, Cohen, and Veličković]{bronstein_geometric_2021}
Bronstein,~M.~M.; Bruna,~J.; Cohen,~T.; Veličković,~P. Geometric {Deep} {Learning}: {Grids}, {Groups}, {Graphs}, {Geodesics}, and {Gauges}. 2021; \url{http://arxiv.org/abs/2104.13478}\relax
\mciteBstWouldAddEndPuncttrue
\mciteSetBstMidEndSepPunct{\mcitedefaultmidpunct}
{\mcitedefaultendpunct}{\mcitedefaultseppunct}\relax
\EndOfBibitem
\bibitem[Reiser \latin{et~al.}(2022)Reiser, Neubert, Eberhard, Torresi, Zhou, Shao, Metni, van Hoesel, Schopmans, Sommer, and Friederich]{reiser_graph_2022}
Reiser,~P.; Neubert,~M.; Eberhard,~A.; Torresi,~L.; Zhou,~C.; Shao,~C.; Metni,~H.; van Hoesel,~C.; Schopmans,~H.; Sommer,~T.; Friederich,~P. Graph neural networks for materials science and chemistry. \emph{Communications Materials} \textbf{2022}, \emph{3}, 93\relax
\mciteBstWouldAddEndPuncttrue
\mciteSetBstMidEndSepPunct{\mcitedefaultmidpunct}
{\mcitedefaultendpunct}{\mcitedefaultseppunct}\relax
\EndOfBibitem
\bibitem[Zhou \latin{et~al.}(2020)Zhou, Cui, Hu, Zhang, Yang, Liu, Wang, Li, and Sun]{zhou_graph_2020}
Zhou,~J.; Cui,~G.; Hu,~S.; Zhang,~Z.; Yang,~C.; Liu,~Z.; Wang,~L.; Li,~C.; Sun,~M. Graph neural networks: {A} review of methods and applications. \emph{AI Open} \textbf{2020}, \emph{1}, 57--81\relax
\mciteBstWouldAddEndPuncttrue
\mciteSetBstMidEndSepPunct{\mcitedefaultmidpunct}
{\mcitedefaultendpunct}{\mcitedefaultseppunct}\relax
\EndOfBibitem
\bibitem[Fung \latin{et~al.}(2021)Fung, Zhang, Juarez, and Sumpter]{fung_benchmarking_2021}
Fung,~V.; Zhang,~J.; Juarez,~E.; Sumpter,~B.~G. Benchmarking graph neural networks for materials chemistry. \emph{npj Computational Materials} \textbf{2021}, \emph{7}, 84\relax
\mciteBstWouldAddEndPuncttrue
\mciteSetBstMidEndSepPunct{\mcitedefaultmidpunct}
{\mcitedefaultendpunct}{\mcitedefaultseppunct}\relax
\EndOfBibitem
\bibitem[Xu \latin{et~al.}(2019)Xu, Hu, Leskovec, and Jegelka]{xu_how_2019}
Xu,~K.; Hu,~W.; Leskovec,~J.; Jegelka,~S. How {Powerful} are {Graph} {Neural} {Networks}? 2019; \url{http://arxiv.org/abs/1810.00826}\relax
\mciteBstWouldAddEndPuncttrue
\mciteSetBstMidEndSepPunct{\mcitedefaultmidpunct}
{\mcitedefaultendpunct}{\mcitedefaultseppunct}\relax
\EndOfBibitem
\bibitem[Xie and Grossman(2018)Xie, and Grossman]{xie_crystal_2018}
Xie,~T.; Grossman,~J.~C. Crystal {Graph} {Convolutional} {Neural} {Networks} for an {Accurate} and {Interpretable} {Prediction} of {Material} {Properties}. \emph{Physical Review Letters} \textbf{2018}, \emph{120}, 145301\relax
\mciteBstWouldAddEndPuncttrue
\mciteSetBstMidEndSepPunct{\mcitedefaultmidpunct}
{\mcitedefaultendpunct}{\mcitedefaultseppunct}\relax
\EndOfBibitem
\bibitem[Zhou \latin{et~al.}(2018)Zhou, Tang, Liu, Pan, Yan, and Zhang]{zhou_atom2vec_2018}
Zhou,~Q.; Tang,~P.; Liu,~S.; Pan,~J.; Yan,~Q.; Zhang,~S.-C. {Atom2Vec}: learning atoms for materials discovery. \emph{Proceedings of the National Academy of Sciences} \textbf{2018}, \emph{115}\relax
\mciteBstWouldAddEndPuncttrue
\mciteSetBstMidEndSepPunct{\mcitedefaultmidpunct}
{\mcitedefaultendpunct}{\mcitedefaultseppunct}\relax
\EndOfBibitem
\bibitem[Thomas \latin{et~al.}(2018)Thomas, Smidt, Kearnes, Yang, Li, Kohlhoff, and Riley]{thomas_tensor_2018}
Thomas,~N.; Smidt,~T.; Kearnes,~S.; Yang,~L.; Li,~L.; Kohlhoff,~K.; Riley,~P. Tensor field networks: {Rotation}- and translation-equivariant neural networks for {3D} point clouds. 2018; \url{http://arxiv.org/abs/1802.08219}\relax
\mciteBstWouldAddEndPuncttrue
\mciteSetBstMidEndSepPunct{\mcitedefaultmidpunct}
{\mcitedefaultendpunct}{\mcitedefaultseppunct}\relax
\EndOfBibitem
\bibitem[Batzner \latin{et~al.}(2022)Batzner, Musaelian, Sun, Geiger, Mailoa, Kornbluth, Molinari, Smidt, and Kozinsky]{batzner_e3-equivariant_2022}
Batzner,~S.; Musaelian,~A.; Sun,~L.; Geiger,~M.; Mailoa,~J.~P.; Kornbluth,~M.; Molinari,~N.; Smidt,~T.~E.; Kozinsky,~B. E(3)-equivariant graph neural networks for data-efficient and accurate interatomic potentials. \emph{Nature Communications} \textbf{2022}, \emph{13}, 1--11\relax
\mciteBstWouldAddEndPuncttrue
\mciteSetBstMidEndSepPunct{\mcitedefaultmidpunct}
{\mcitedefaultendpunct}{\mcitedefaultseppunct}\relax
\EndOfBibitem
\bibitem[Liao \latin{et~al.}(2024)Liao, Wood, Das, and Smidt]{liao_equiformerv2_2024}
Liao,~Y.-L.; Wood,~B.; Das,~A.; Smidt,~T. {EquiformerV2}: {Improved} {Equivariant} {Transformer} for {Scaling} to {Higher}-{Degree} {Representations}. 2024; \url{http://arxiv.org/abs/2306.12059}\relax
\mciteBstWouldAddEndPuncttrue
\mciteSetBstMidEndSepPunct{\mcitedefaultmidpunct}
{\mcitedefaultendpunct}{\mcitedefaultseppunct}\relax
\EndOfBibitem
\bibitem[Freeman and Adelson(1991)Freeman, and Adelson]{freeman_design_1991}
Freeman,~W.; Adelson,~E. The design and use of steerable filters. \emph{IEEE Transactions on Pattern Analysis and Machine Intelligence} \textbf{1991}, \emph{13}, 891--906\relax
\mciteBstWouldAddEndPuncttrue
\mciteSetBstMidEndSepPunct{\mcitedefaultmidpunct}
{\mcitedefaultendpunct}{\mcitedefaultseppunct}\relax
\EndOfBibitem
\bibitem[Cohen and Welling(2016)Cohen, and Welling]{cohen_steerable_2016}
Cohen,~T.~S.; Welling,~M. Steerable {CNNs}. 2016; \url{http://arxiv.org/abs/1612.08498}\relax
\mciteBstWouldAddEndPuncttrue
\mciteSetBstMidEndSepPunct{\mcitedefaultmidpunct}
{\mcitedefaultendpunct}{\mcitedefaultseppunct}\relax
\EndOfBibitem
\bibitem[Worrall \latin{et~al.}(2017)Worrall, Garbin, Turmukhambetov, and Brostow]{worrall_harmonic_2017}
Worrall,~D.~E.; Garbin,~S.~J.; Turmukhambetov,~D.; Brostow,~G.~J. Harmonic {Networks}: {Deep} {Translation} and {Rotation} {Equivariance}. 2017; \url{http://arxiv.org/abs/1612.04642}\relax
\mciteBstWouldAddEndPuncttrue
\mciteSetBstMidEndSepPunct{\mcitedefaultmidpunct}
{\mcitedefaultendpunct}{\mcitedefaultseppunct}\relax
\EndOfBibitem
\bibitem[Schütt \latin{et~al.}(2018)Schütt, Sauceda, Kindermans, Tkatchenko, and Müller]{schutt_schnet_2018}
Schütt,~K.~T.; Sauceda,~H.~E.; Kindermans,~P.-J.; Tkatchenko,~A.; Müller,~K.-R. {SchNet} – {A} deep learning architecture for molecules and materials. \emph{The Journal of Chemical Physics} \textbf{2018}, \emph{148}, 241722\relax
\mciteBstWouldAddEndPuncttrue
\mciteSetBstMidEndSepPunct{\mcitedefaultmidpunct}
{\mcitedefaultendpunct}{\mcitedefaultseppunct}\relax
\EndOfBibitem
\bibitem[Wen \latin{et~al.}(2024)Wen, K. Horton, M. Munro, Huck, and A. Persson]{wen_equivariant_2024}
Wen,~M.; K. Horton,~M.; M. Munro,~J.; Huck,~P.; A. Persson,~K. An equivariant graph neural network for the elasticity tensors of all seven crystal systems. \emph{Digital Discovery} \textbf{2024}, \relax
\mciteBstWouldAddEndPunctfalse
\mciteSetBstMidEndSepPunct{\mcitedefaultmidpunct}
{}{\mcitedefaultseppunct}\relax
\EndOfBibitem
\bibitem[Chen \latin{et~al.}(2020)Chen, Kim, Batra, Lightstone, Wu, Li, Deshmukh, Wang, Tran, Vashishta, Sotzing, Cao, and Ramprasad]{chen_frequency-dependent_2020}
Chen,~L.; Kim,~C.; Batra,~R.; Lightstone,~J.~P.; Wu,~C.; Li,~Z.; Deshmukh,~A.~A.; Wang,~Y.; Tran,~H.~D.; Vashishta,~P.; Sotzing,~G.~A.; Cao,~Y.; Ramprasad,~R. Frequency-dependent dielectric constant prediction of polymers using machine learning. \emph{npj Computational Materials} \textbf{2020}, \emph{6}, 61\relax
\mciteBstWouldAddEndPuncttrue
\mciteSetBstMidEndSepPunct{\mcitedefaultmidpunct}
{\mcitedefaultendpunct}{\mcitedefaultseppunct}\relax
\EndOfBibitem
\bibitem[Morita \latin{et~al.}(2020)Morita, Davies, Butler, and Walsh]{morita_modeling_2020}
Morita,~K.; Davies,~D.~W.; Butler,~K.~T.; Walsh,~A. Modeling the dielectric constants of crystals using machine learning. \emph{The Journal of Chemical Physics} \textbf{2020}, \emph{153}, 024503\relax
\mciteBstWouldAddEndPuncttrue
\mciteSetBstMidEndSepPunct{\mcitedefaultmidpunct}
{\mcitedefaultendpunct}{\mcitedefaultseppunct}\relax
\EndOfBibitem
\bibitem[Takahashi \latin{et~al.}(2020)Takahashi, Kumagai, Miyamoto, Mochizuki, and Oba]{takahashi_machine_2020}
Takahashi,~A.; Kumagai,~Y.; Miyamoto,~J.; Mochizuki,~Y.; Oba,~F. Machine learning models for predicting the dielectric constants of oxides based on high-throughput first-principles calculations. \emph{Physical Review Materials} \textbf{2020}, \emph{4}, 103801\relax
\mciteBstWouldAddEndPuncttrue
\mciteSetBstMidEndSepPunct{\mcitedefaultmidpunct}
{\mcitedefaultendpunct}{\mcitedefaultseppunct}\relax
\EndOfBibitem
\bibitem[Grisafi \latin{et~al.}(2018)Grisafi, Wilkins, Csányi, and Ceriotti]{grisafi_symmetry-adapted_2018}
Grisafi,~A.; Wilkins,~D.~M.; Csányi,~G.; Ceriotti,~M. Symmetry-{Adapted} {Machine} {Learning} for {Tensorial} {Properties} of {Atomistic} {Systems}. \emph{Physical Review Letters} \textbf{2018}, \emph{120}, 036002\relax
\mciteBstWouldAddEndPuncttrue
\mciteSetBstMidEndSepPunct{\mcitedefaultmidpunct}
{\mcitedefaultendpunct}{\mcitedefaultseppunct}\relax
\EndOfBibitem
\bibitem[Falletta \latin{et~al.}(2024)Falletta, Cepellotti, Tan, Johansson, Musaelian, Owen, and Kozinsky]{falletta2024unified}
Falletta,~S.; Cepellotti,~A.; Tan,~C.~W.; Johansson,~A.; Musaelian,~A.; Owen,~C.~J.; Kozinsky,~B. Unified Differentiable Learning of the Electric Enthalpy and Dielectric Properties with Exact Physical Constraints. 2024; \url{https://doi.org/10.48550/arXiv.2403.17207}\relax
\mciteBstWouldAddEndPuncttrue
\mciteSetBstMidEndSepPunct{\mcitedefaultmidpunct}
{\mcitedefaultendpunct}{\mcitedefaultseppunct}\relax
\EndOfBibitem
\bibitem[Prati(2003)]{prati_propagation_2003}
Prati,~E. Propagation in {Gyroelectromagnetic} {Guiding} {Systems}. \emph{Journal of Electromagnetic Waves and Applications} \textbf{2003}, \emph{17}, 1177--1196\relax
\mciteBstWouldAddEndPuncttrue
\mciteSetBstMidEndSepPunct{\mcitedefaultmidpunct}
{\mcitedefaultendpunct}{\mcitedefaultseppunct}\relax
\EndOfBibitem
\bibitem[Jahani and Jacob(2016)Jahani, and Jacob]{jahani_all-dielectric_2016}
Jahani,~S.; Jacob,~Z. All-dielectric metamaterials. \emph{Nature Nanotechnology} \textbf{2016}, \emph{11}, 23--36\relax
\mciteBstWouldAddEndPuncttrue
\mciteSetBstMidEndSepPunct{\mcitedefaultmidpunct}
{\mcitedefaultendpunct}{\mcitedefaultseppunct}\relax
\EndOfBibitem
\bibitem[Xie \latin{et~al.}(2022)Xie, Shi, Feng, Sun, Liu, Yan, Liu, Moussa, Huang, Meng, Liang, Hou, Fan, and Guo]{xie_recent_2022}
Xie,~P.; Shi,~Z.; Feng,~M.; Sun,~K.; Liu,~Y.; Yan,~K.; Liu,~C.; Moussa,~T. A.~A.; Huang,~M.; Meng,~S.; Liang,~G.; Hou,~H.; Fan,~R.; Guo,~Z. Recent advances in radio-frequency negative dielectric metamaterials by designing heterogeneous composites. \emph{Advanced Composites and Hybrid Materials} \textbf{2022}, \emph{5}, 679--695\relax
\mciteBstWouldAddEndPuncttrue
\mciteSetBstMidEndSepPunct{\mcitedefaultmidpunct}
{\mcitedefaultendpunct}{\mcitedefaultseppunct}\relax
\EndOfBibitem
\bibitem[Lee \latin{et~al.}(2018)Lee, Youn, Yim, and Han]{lee_high-throughput_2018}
Lee,~M.; Youn,~Y.; Yim,~K.; Han,~S. High-throughput ab initio calculations on dielectric constant and band gap of non-oxide dielectrics. \emph{Scientific Reports} \textbf{2018}, \emph{8}, 14794\relax
\mciteBstWouldAddEndPuncttrue
\mciteSetBstMidEndSepPunct{\mcitedefaultmidpunct}
{\mcitedefaultendpunct}{\mcitedefaultseppunct}\relax
\EndOfBibitem
\bibitem[Ong \latin{et~al.}(2015)Ong, Cholia, Jain, Brafman, Gunter, Ceder, and Persson]{ong_materials_2015}
Ong,~S.~P.; Cholia,~S.; Jain,~A.; Brafman,~M.; Gunter,~D.; Ceder,~G.; Persson,~K.~A. The {Materials} {Application} {Programming} {Interface} ({API}): {A} simple, flexible and efficient {API} for materials data based on {REpresentational} {State} {Transfer} ({REST}) principles. \emph{Computational Materials Science} \textbf{2015}, \emph{97}, 209--215\relax
\mciteBstWouldAddEndPuncttrue
\mciteSetBstMidEndSepPunct{\mcitedefaultmidpunct}
{\mcitedefaultendpunct}{\mcitedefaultseppunct}\relax
\EndOfBibitem
\bibitem[Perdew \latin{et~al.}(1996)Perdew, Burke, and Ernzerhof]{perdew_generalized_1996}
Perdew,~J.~P.; Burke,~K.; Ernzerhof,~M. Generalized {Gradient} {Approximation} {Made} {Simple}. \emph{Physical Review Letters} \textbf{1996}, \emph{77}, 3865--3868\relax
\mciteBstWouldAddEndPuncttrue
\mciteSetBstMidEndSepPunct{\mcitedefaultmidpunct}
{\mcitedefaultendpunct}{\mcitedefaultseppunct}\relax
\EndOfBibitem
\bibitem[Geiger and Smidt(2022)Geiger, and Smidt]{geiger_e3nn_2022}
Geiger,~M.; Smidt,~T. e3nn: {Euclidean} {Neural} {Networks}. 2022; \url{http://arxiv.org/abs/2207.09453}\relax
\mciteBstWouldAddEndPuncttrue
\mciteSetBstMidEndSepPunct{\mcitedefaultmidpunct}
{\mcitedefaultendpunct}{\mcitedefaultseppunct}\relax
\EndOfBibitem
\bibitem[Paszke \latin{et~al.}(2017)Paszke, Gross, Chintala, Chanan, Yang, DeVito, Lin, Desmaison, Antiga, and Lerer]{paszke2017automatic}
Paszke,~A.; Gross,~S.; Chintala,~S.; Chanan,~G.; Yang,~E.; DeVito,~Z.; Lin,~Z.; Desmaison,~A.; Antiga,~L.; Lerer,~A. Automatic differentiation in PyTorch. \textbf{2017}, \relax
\mciteBstWouldAddEndPunctfalse
\mciteSetBstMidEndSepPunct{\mcitedefaultmidpunct}
{}{\mcitedefaultseppunct}\relax
\EndOfBibitem
\bibitem[Jørgensen and Bhowmik(2022)Jørgensen, and Bhowmik]{jorgensen_equivariant_2022}
Jørgensen,~P.~B.; Bhowmik,~A. Equivariant graph neural networks for fast electron density estimation of molecules, liquids, and solids. \emph{npj Computational Materials} \textbf{2022}, \emph{8}, 1--10\relax
\mciteBstWouldAddEndPuncttrue
\mciteSetBstMidEndSepPunct{\mcitedefaultmidpunct}
{\mcitedefaultendpunct}{\mcitedefaultseppunct}\relax
\EndOfBibitem
\bibitem[Batatia \latin{et~al.}(2023)Batatia, Kovács, Simm, Ortner, and Csányi]{batatia_mace_2023}
Batatia,~I.; Kovács,~D.~P.; Simm,~G. N.~C.; Ortner,~C.; Csányi,~G. {MACE}: {Higher} {Order} {Equivariant} {Message} {Passing} {Neural} {Networks} for {Fast} and {Accurate} {Force} {Fields}. 2023; \url{http://arxiv.org/abs/2206.07697}\relax
\mciteBstWouldAddEndPuncttrue
\mciteSetBstMidEndSepPunct{\mcitedefaultmidpunct}
{\mcitedefaultendpunct}{\mcitedefaultseppunct}\relax
\EndOfBibitem
\bibitem[Chen \latin{et~al.}(2021)Chen, Andrejevic, Smidt, Ding, Xu, Chi, Nguyen, Alatas, Kong, and Li]{chen_phonon_2019}
Chen,~Z.; Andrejevic,~N.; Smidt,~T.; Ding,~Z.; Xu,~Q.; Chi,~Y.-T.; Nguyen,~Q.~T.; Alatas,~A.; Kong,~J.; Li,~M. Direct Prediction of Phonon Density of States With Euclidean Neural Networks. \emph{Advanced Science} \textbf{2021}, \emph{8}, 2004214\relax
\mciteBstWouldAddEndPuncttrue
\mciteSetBstMidEndSepPunct{\mcitedefaultmidpunct}
{\mcitedefaultendpunct}{\mcitedefaultseppunct}\relax
\EndOfBibitem
\bibitem[Chen \latin{et~al.}(2019)Chen, Ye, Zuo, Zheng, and Ong]{chen_graph_2019}
Chen,~C.; Ye,~W.; Zuo,~Y.; Zheng,~C.; Ong,~S.~P. Graph {Networks} as a {Universal} {Machine} {Learning} {Framework} for {Molecules} and {Crystals}. \emph{Chemistry of Materials} \textbf{2019}, \emph{31}, 3564--3572\relax
\mciteBstWouldAddEndPuncttrue
\mciteSetBstMidEndSepPunct{\mcitedefaultmidpunct}
{\mcitedefaultendpunct}{\mcitedefaultseppunct}\relax
\EndOfBibitem
\bibitem[Chen and Ong(2022)Chen, and Ong]{chen_universal_2022}
Chen,~C.; Ong,~S.~P. A {Universal} {Graph} {Deep} {Learning} {Interatomic} {Potential} for the {Periodic} {Table}. \emph{Nature Computational Science} \textbf{2022}, \emph{2}, 718--728\relax
\mciteBstWouldAddEndPuncttrue
\mciteSetBstMidEndSepPunct{\mcitedefaultmidpunct}
{\mcitedefaultendpunct}{\mcitedefaultseppunct}\relax
\EndOfBibitem
\bibitem[Sanyal \latin{et~al.}(2018)Sanyal, Balachandran, Yadati, Kumar, Rajagopalan, Sanyal, and Talukdar]{sanyal_mt-cgcnn_2018}
Sanyal,~S.; Balachandran,~J.; Yadati,~N.; Kumar,~A.; Rajagopalan,~P.; Sanyal,~S.; Talukdar,~P. {MT}-{CGCNN}: {Integrating} {Crystal} {Graph} {Convolutional} {Neural} {Network} with {Multitask} {Learning} for {Material} {Property} {Prediction}. 2018; \url{http://arxiv.org/abs/1811.05660}\relax
\mciteBstWouldAddEndPuncttrue
\mciteSetBstMidEndSepPunct{\mcitedefaultmidpunct}
{\mcitedefaultendpunct}{\mcitedefaultseppunct}\relax
\EndOfBibitem
\bibitem[Backus(1970)]{backus_geometrical_1970}
Backus,~G. A geometrical picture of anisotropic elastic tensors. \emph{Reviews of Geophysics} \textbf{1970}, \emph{8}, 633--671\relax
\mciteBstWouldAddEndPuncttrue
\mciteSetBstMidEndSepPunct{\mcitedefaultmidpunct}
{\mcitedefaultendpunct}{\mcitedefaultseppunct}\relax
\EndOfBibitem
\bibitem[Smidt \latin{et~al.}(2021)Smidt, Geiger, and Miller]{smidt_finding_2021}
Smidt,~T.~E.; Geiger,~M.; Miller,~B.~K. Finding symmetry breaking order parameters with {Euclidean} neural networks. \emph{Physical Review Research} \textbf{2021}, \emph{3}, L012002\relax
\mciteBstWouldAddEndPuncttrue
\mciteSetBstMidEndSepPunct{\mcitedefaultmidpunct}
{\mcitedefaultendpunct}{\mcitedefaultseppunct}\relax
\EndOfBibitem
\bibitem[Loshchilov and Hutter(2019)Loshchilov, and Hutter]{loshchilov_decoupled_2019}
Loshchilov,~I.; Hutter,~F. Decoupled {Weight} {Decay} {Regularization}. 2019; \url{http://arxiv.org/abs/1711.05101}\relax
\mciteBstWouldAddEndPuncttrue
\mciteSetBstMidEndSepPunct{\mcitedefaultmidpunct}
{\mcitedefaultendpunct}{\mcitedefaultseppunct}\relax
\EndOfBibitem
\bibitem[Akiba \latin{et~al.}(2019)Akiba, Sano, Yanase, Ohta, and Koyama]{akiba_optuna_2019}
Akiba,~T.; Sano,~S.; Yanase,~T.; Ohta,~T.; Koyama,~M. Optuna: {A} {Next}-generation {Hyperparameter} {Optimization} {Framework}. Proceedings of the 25th {ACM} {SIGKDD} {International} {Conference} on {Knowledge} {Discovery} \& {Data} {Mining}. New York, NY, USA, 2019; pp 2623--2631\relax
\mciteBstWouldAddEndPuncttrue
\mciteSetBstMidEndSepPunct{\mcitedefaultmidpunct}
{\mcitedefaultendpunct}{\mcitedefaultseppunct}\relax
\EndOfBibitem
\bibitem[Kresse and Hafner(1993)Kresse, and Hafner]{kresse_initio_1993a}
Kresse,~G.; Hafner,~J. Ab Initio Molecular Dynamics for Liquid Metals. \emph{Physical Review B} \textbf{1993}, \emph{47}, 558--561\relax
\mciteBstWouldAddEndPuncttrue
\mciteSetBstMidEndSepPunct{\mcitedefaultmidpunct}
{\mcitedefaultendpunct}{\mcitedefaultseppunct}\relax
\EndOfBibitem
\bibitem[Kresse and Furthm{\"u}ller(1996)Kresse, and Furthm{\"u}ller]{kresse_efficiency_1996}
Kresse,~G.; Furthm{\"u}ller,~J. Efficiency of Ab-Initio Total Energy Calculations for Metals and Semiconductors Using a Plane-Wave Basis Set. \emph{Computational Materials Science} \textbf{1996}, \emph{6}, 15--50\relax
\mciteBstWouldAddEndPuncttrue
\mciteSetBstMidEndSepPunct{\mcitedefaultmidpunct}
{\mcitedefaultendpunct}{\mcitedefaultseppunct}\relax
\EndOfBibitem
\bibitem[Ganose \latin{et~al.}(2024)Ganose, Riebesell, George, Shen, S.~Rosen, Ashok~Naik, Winner, Wen, Guha, Kuner, Petretto, Zhu, Horton, Sahasrabuddhe, Kaplan, Schmidt, Ertural, Kingsbury, McDermott, Goodall, Bonkowski, Purcell, Z\"{u}gner, and Qi]{Ganose_atomate2_2024}
Ganose,~A. \latin{et~al.}  {atomate2}. 2024; \url{https://github.com/materialsproject/atomate2}\relax
\mciteBstWouldAddEndPuncttrue
\mciteSetBstMidEndSepPunct{\mcitedefaultmidpunct}
{\mcitedefaultendpunct}{\mcitedefaultseppunct}\relax
\EndOfBibitem
\bibitem[Rosen \latin{et~al.}(2024)Rosen, Gallant, George, Riebesell, Sahasrabuddhe, Shen, Wen, Evans, Petretto, Waroquiers, Rignanese, Persson, Jain, and Ganose]{Rosen_Jobflow_Computational_Workflows_2024}
Rosen,~A.~S.; Gallant,~M.; George,~J.; Riebesell,~J.; Sahasrabuddhe,~H.; Shen,~J.-X.; Wen,~M.; Evans,~M.~L.; Petretto,~G.; Waroquiers,~D.; Rignanese,~G.-M.; Persson,~K.~A.; Jain,~A.; Ganose,~A.~M. {Jobflow: Computational Workflows Made Simple}. \emph{Journal of Open Source Software} \textbf{2024}, \emph{9}, 5995\relax
\mciteBstWouldAddEndPuncttrue
\mciteSetBstMidEndSepPunct{\mcitedefaultmidpunct}
{\mcitedefaultendpunct}{\mcitedefaultseppunct}\relax
\EndOfBibitem
\bibitem[job(2024)]{jobflow_remote}
jobflow-remote. \url{https://github.com/Matgenix/jobflow-remote}, 2024\relax
\mciteBstWouldAddEndPuncttrue
\mciteSetBstMidEndSepPunct{\mcitedefaultmidpunct}
{\mcitedefaultendpunct}{\mcitedefaultseppunct}\relax
\EndOfBibitem
\bibitem[Ong \latin{et~al.}(2013)Ong, Richards, Jain, Hautier, Kocher, Cholia, Gunter, Chevrier, Persson, and Ceder]{ong_python_2013}
Ong,~S.~P.; Richards,~W.~D.; Jain,~A.; Hautier,~G.; Kocher,~M.; Cholia,~S.; Gunter,~D.; Chevrier,~V.~L.; Persson,~K.~A.; Ceder,~G. Python {Materials} {Genomics} (pymatgen): {A} robust, open-source python library for materials analysis. \emph{Computational Materials Science} \textbf{2013}, \emph{68}, 314--319\relax
\mciteBstWouldAddEndPuncttrue
\mciteSetBstMidEndSepPunct{\mcitedefaultmidpunct}
{\mcitedefaultendpunct}{\mcitedefaultseppunct}\relax
\EndOfBibitem
\bibitem[Gajdo{\v s} \latin{et~al.}(2006)Gajdo{\v s}, Hummer, Kresse, Furthm{\"u}ller, and Bechstedt]{gajdos_linear_2006}
Gajdo{\v s},~M.; Hummer,~K.; Kresse,~G.; Furthm{\"u}ller,~J.; Bechstedt,~F. Linear Optical Properties in the Projector-Augmented Wave Methodology. \emph{Physical Review B} \textbf{2006}, \emph{73}, 045112\relax
\mciteBstWouldAddEndPuncttrue
\mciteSetBstMidEndSepPunct{\mcitedefaultmidpunct}
{\mcitedefaultendpunct}{\mcitedefaultseppunct}\relax
\EndOfBibitem
\bibitem[Petousis \latin{et~al.}(2016)Petousis, Chen, Hautier, Graf, Schladt, Persson, and Prinz]{petousis2016benchmarking}
Petousis,~I.; Chen,~W.; Hautier,~G.; Graf,~T.; Schladt,~T.~D.; Persson,~K.~A.; Prinz,~F.~B. Benchmarking density functional perturbation theory to enable high-throughput screening of materials for dielectric constant and refractive index. \emph{Physical Review B} \textbf{2016}, \emph{93}, 115151\relax
\mciteBstWouldAddEndPuncttrue
\mciteSetBstMidEndSepPunct{\mcitedefaultmidpunct}
{\mcitedefaultendpunct}{\mcitedefaultseppunct}\relax
\EndOfBibitem
\bibitem[Dunn \latin{et~al.}(2020)Dunn, Wang, Ganose, Dopp, and Jain]{dunn_benchmarking_2020}
Dunn,~A.; Wang,~Q.; Ganose,~A.; Dopp,~D.; Jain,~A. Benchmarking materials property prediction methods: the {Matbench} test set and {Automatminer} reference algorithm. \emph{npj Computational Materials} \textbf{2020}, \emph{6}, 1--10\relax
\mciteBstWouldAddEndPuncttrue
\mciteSetBstMidEndSepPunct{\mcitedefaultmidpunct}
{\mcitedefaultendpunct}{\mcitedefaultseppunct}\relax
\EndOfBibitem
\bibitem[Ruff \latin{et~al.}(2024)Ruff, Reiser, Stühmer, and Friederich]{ruff_connectivity_2024}
Ruff,~R.; Reiser,~P.; Stühmer,~J.; Friederich,~P. Connectivity optimized nested line graph networks for crystal structures. \emph{Digital Discovery} \textbf{2024}, \emph{3}, 594--601\relax
\mciteBstWouldAddEndPuncttrue
\mciteSetBstMidEndSepPunct{\mcitedefaultmidpunct}
{\mcitedefaultendpunct}{\mcitedefaultseppunct}\relax
\EndOfBibitem
\bibitem[Neumann and Meyer(1885)Neumann, and Meyer]{neumann1885vorlesungen}
Neumann,~F.~L.; Meyer,~O.~E. \emph{Vorlesungen über die theorie der elasticität der festen körper und des lichtäthers}; Druck und Verlag von BG Teubner, 1885\relax
\mciteBstWouldAddEndPuncttrue
\mciteSetBstMidEndSepPunct{\mcitedefaultmidpunct}
{\mcitedefaultendpunct}{\mcitedefaultseppunct}\relax
\EndOfBibitem
\bibitem[Tudi \latin{et~al.}(2022)Tudi, Han, Yang, and Pan]{tudi2022potential}
Tudi,~A.; Han,~S.; Yang,~Z.; Pan,~S. Potential optical functional crystals with large birefringence: Recent advances and future prospects. \emph{Coordination Chemistry Reviews} \textbf{2022}, \emph{459}, 214380\relax
\mciteBstWouldAddEndPuncttrue
\mciteSetBstMidEndSepPunct{\mcitedefaultmidpunct}
{\mcitedefaultendpunct}{\mcitedefaultseppunct}\relax
\EndOfBibitem
\bibitem[Liu \latin{et~al.}(2021)Liu, Laguta, Inzani, Huang, Das, Chatterjee, Sheridan, Griffin, Ardavan, and Ramesh]{liu2021coherent}
Liu,~J.; Laguta,~V.~V.; Inzani,~K.; Huang,~W.; Das,~S.; Chatterjee,~R.; Sheridan,~E.; Griffin,~S.~M.; Ardavan,~A.; Ramesh,~R. Coherent electric field manipulation of Fe3+ spins in PbTiO3. \emph{Science Advances} \textbf{2021}, \emph{7}, eabf8103\relax
\mciteBstWouldAddEndPuncttrue
\mciteSetBstMidEndSepPunct{\mcitedefaultmidpunct}
{\mcitedefaultendpunct}{\mcitedefaultseppunct}\relax
\EndOfBibitem
\bibitem[Ganose and Jain(2019)Ganose, and Jain]{Ganose_Jain_2019}
Ganose,~A.~M.; Jain,~A. Robocrystallographer: automated crystal structure text descriptions and analysis. \emph{MRS Communications} \textbf{2019}, \emph{9}, 874–881\relax
\mciteBstWouldAddEndPuncttrue
\mciteSetBstMidEndSepPunct{\mcitedefaultmidpunct}
{\mcitedefaultendpunct}{\mcitedefaultseppunct}\relax
\EndOfBibitem
\bibitem[Pan \latin{et~al.}(2021)Pan, Ganose, Horton, Aykol, Persson, Zimmermann, and Jain]{pan2021benchmarking}
Pan,~H.; Ganose,~A.~M.; Horton,~M.; Aykol,~M.; Persson,~K.~A.; Zimmermann,~N.~E.; Jain,~A. Benchmarking coordination number prediction algorithms on inorganic crystal structures. \emph{Inorganic chemistry} \textbf{2021}, \emph{60}, 1590--1603\relax
\mciteBstWouldAddEndPuncttrue
\mciteSetBstMidEndSepPunct{\mcitedefaultmidpunct}
{\mcitedefaultendpunct}{\mcitedefaultseppunct}\relax
\EndOfBibitem
\bibitem[Di~Giovanni \latin{et~al.}(2023)Di~Giovanni, Giusti, Barbero, Luise, Lio', and Bronstein]{di_giovanni_over-squashing_2023}
Di~Giovanni,~F.; Giusti,~L.; Barbero,~F.; Luise,~G.; Lio',~P.; Bronstein,~M. On {Over}-{Squashing} in {Message} {Passing} {Neural} {Networks}: {The} {Impact} of {Width}, {Depth}, and {Topology}. 2023; \url{http://arxiv.org/abs/2302.02941}\relax
\mciteBstWouldAddEndPuncttrue
\mciteSetBstMidEndSepPunct{\mcitedefaultmidpunct}
{\mcitedefaultendpunct}{\mcitedefaultseppunct}\relax
\EndOfBibitem
\end{mcitethebibliography}

\end{document}


\section{AnisoNet performance on train and test splits}

\begin{figure}[H]
    \centering
    \includegraphics[width=\textwidth]{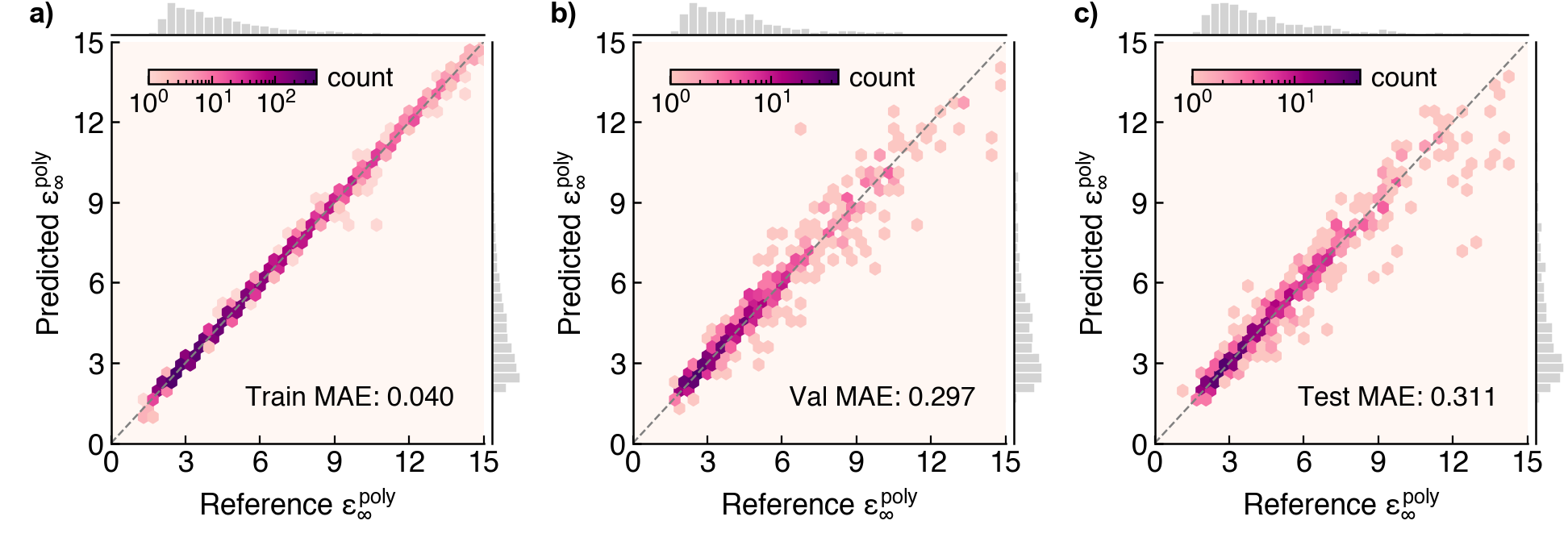}
    \caption{Performance of AnisoNet at predicting the polycrystalline dielectric constant, $\epsilon_\infty^\mathrm{poly}$ on the Materials Project dielectric a) train, b) validation, and c) test sets. The data is presented as heatmaps of reference vs predicted dielectric constants where a darker colour indicates more materials.}
\end{figure}

\begin{figure}[H]
    \centering
    \includegraphics[width=\textwidth]{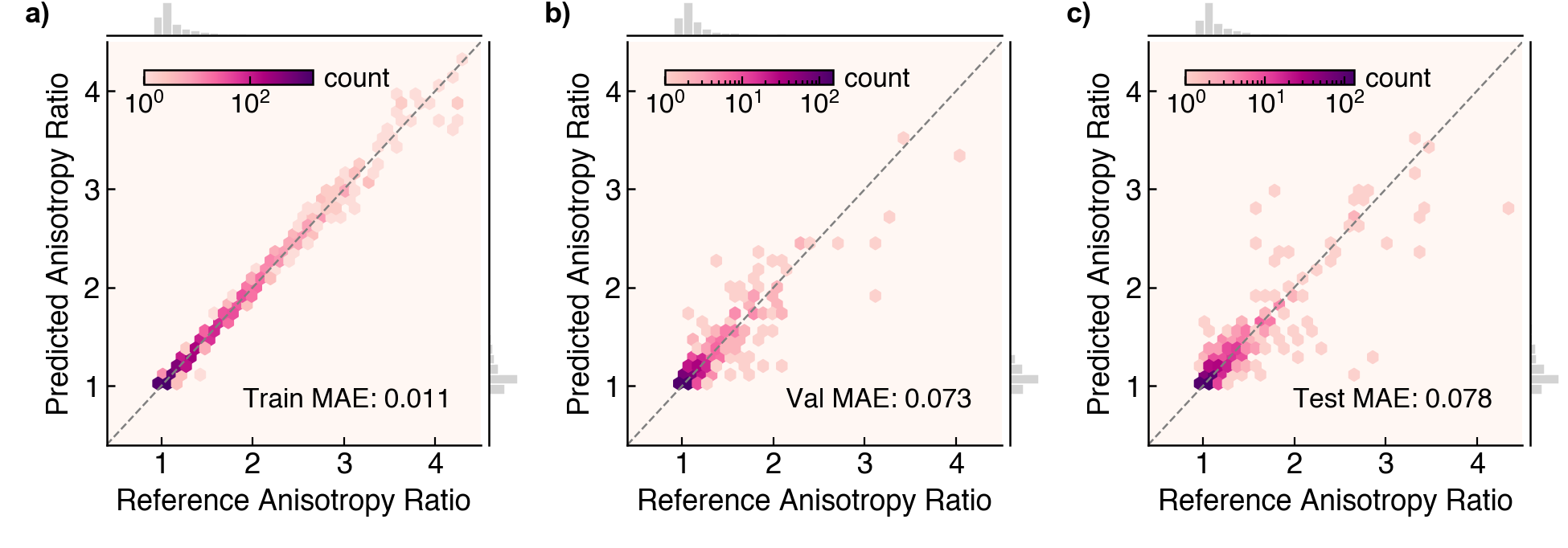}
    \caption{Performance of AnisoNet at predicting the anisotropy ratio on the Materials Project dielectric a) train, b) validation, and c) test sets. The data is presented as heatmaps of reference vs predicted dielectric constants where a darker colour indicates more materials.}
\end{figure}

\section{Analysis of predicted Materials Project dataset}

While almost all materials have a predicted $\varepsilon_\infty^\mathrm{poly}$ below 15 (the cut-off for the training set), several outliers exhibited values up to 30. The validity of these predictions should be speculated since there is no material with a  higher than 15 for the model to learn from during training. The predicted anisotropy for the Materials Project dataset peaks at 1, with most predictions lying between 1 to 3 and outliers with an $a_\mathrm{r}$ up to 9. 

\begin{figure}
    \centering
    \includegraphics[width=0.6\textwidth]{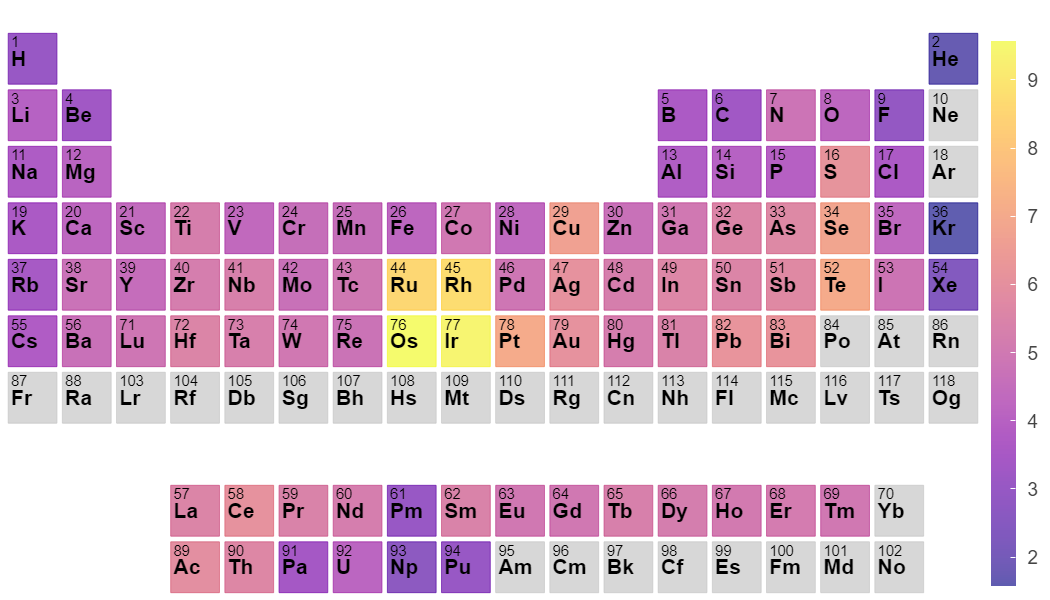}
    \caption{Average predicted polycrystalline dielectric constant ($\epsilon_\infty^\mathrm{poly}$) of the Materials Project filtered dataset (MP-filtered) across the periodic table. Each element's colour corresponds to the average $\epsilon_\infty^\mathrm{poly}$ for all compounds containing that element, and a brighter colour indicates a higher $\epsilon_\infty^\mathrm{poly}$.}
\end{figure}

\begin{figure}
    \centering
    \includegraphics[width=0.6\textwidth]{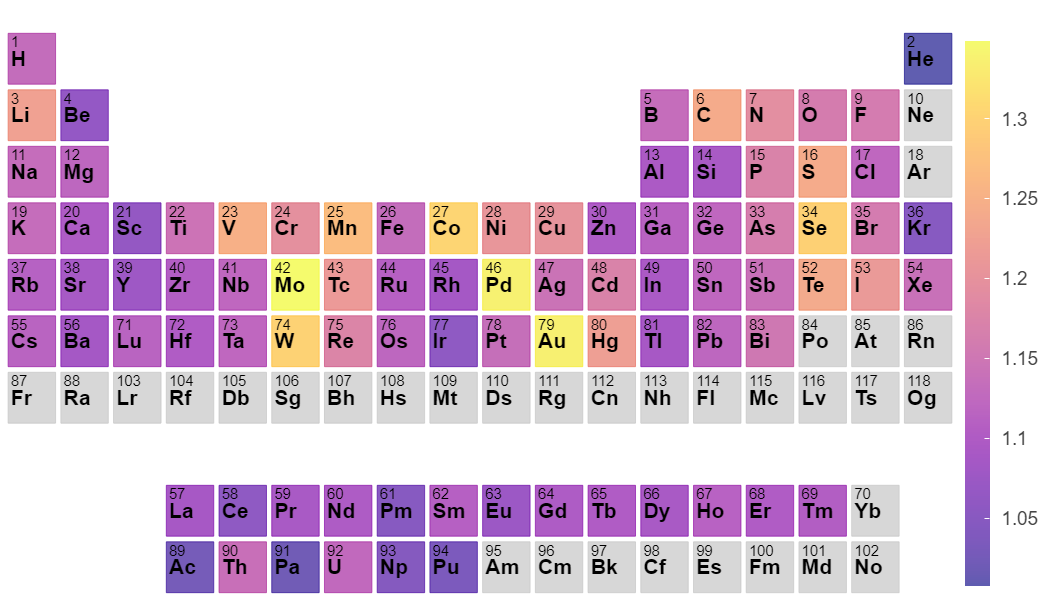}
    \caption{Average anisotropy ratio ($a_\mathrm{r}$) for the MP-filtered dataset across the periodic table. Brighter colour indicates higher $a_\mathrm{r}$.}
\end{figure}

\begin{figure}
    \centering
    \includegraphics[width=0.6\textwidth]{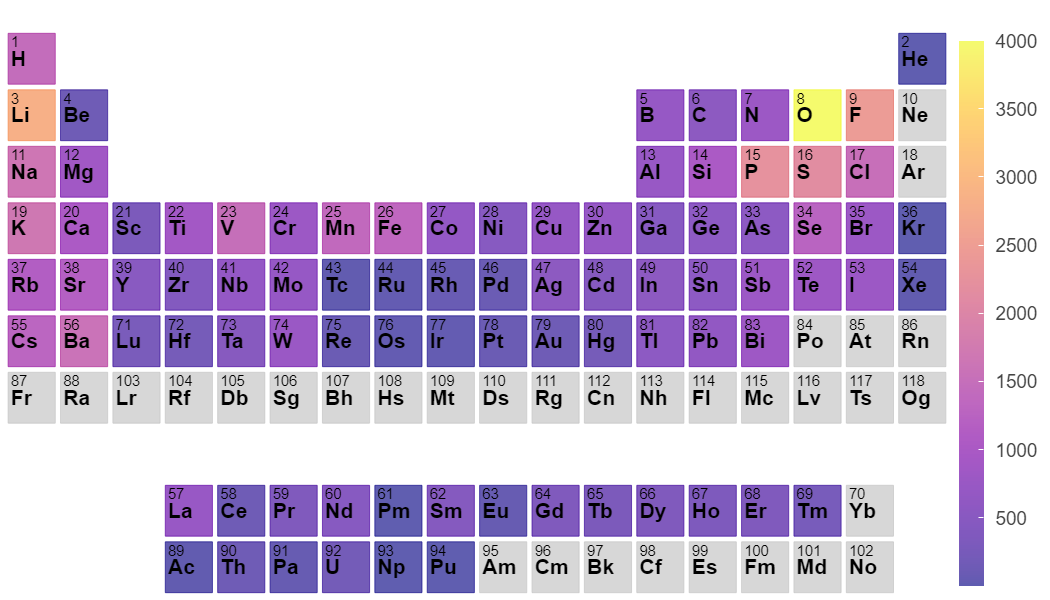}
    \caption{Number of compounds containing a certain element in the MP-filtered dataset. Note that there are 11,990 materials containing oxygen, but the colourbar is capped at 4,000 to increase the contrast between other elements.}
\end{figure}

\clearpage

\section{Anisotropic materials discovery}

\begin{longtable}{lllc}
\caption{List of materials with the highest predicted anisotropy ratio, $a_\mathrm{r}$, and their DFPT-calculated dielectric tensors. \textit{Ab initio} calculations were performed using the Vienna \textit{ab initio} Simulation Package (VASP)\cite{kresse_initio_1993a,kresse_efficiency_1996} and executed using the \textsc{jobflow}\cite{Rosen_Jobflow_Computational_Workflows_2024} and \textsc{jobflow-remote} libraries.} \\
\toprule
Material ID & Formula & Dielectric Tensor & Anisotropy Ratio \\
\midrule
\addlinespace[2ex]
mp-1104369 & \ce{NaV2O4} & $\begin{bmatrix}
2.7 & 0.0 & -0.3 \\ 
0.0 & 18.4 & 0.0 \\ 
-0.3 & 0.0 & 20.8 \\ 
\end{bmatrix}$ & 7.74 \\
\addlinespace[2ex]
mp-1313454 & \ce{LiMnCo3O8} & $\begin{bmatrix}
15.0 & -1.3 & 0.0 \\ 
-1.3 & 10.2 & 0.0 \\ 
0.0 & 0.0 & 3.1 \\ 
\end{bmatrix}$ & 5.94 \\
\addlinespace[2ex]
mp-1023938 & \ce{Te2Mo} & $\begin{bmatrix}
11.5 & 0.0 & 0.0 \\ 
0.0 & 11.5 & 0.0 \\ 
0.0 & 0.0 & 2.0 \\ 
\end{bmatrix}$ & 5.63 \\
\addlinespace[2ex]
mp-1025649 & \ce{Te6Mo2W} & $\begin{bmatrix}
12.9 & 0.0 & 0.0 \\ 
0.0 & 12.9 & 0.0 \\ 
0.0 & 0.0 & 2.4 \\ 
\end{bmatrix}$ & 5.28 \\
\addlinespace[2ex]
mp-1025629 & \ce{Te6Mo2W} & $\begin{bmatrix}
13.1 & 0.0 & 0.0 \\ 
0.0 & 13.1 & 0.0 \\ 
0.0 & 0.0 & 2.5 \\ 
\end{bmatrix}$ & 5.24 \\
\addlinespace[2ex]
mp-1026351 & \ce{Te6MoW2} & $\begin{bmatrix}
12.6 & 0.0 & 0.0 \\ 
0.0 & 12.6 & 0.0 \\ 
0.0 & 0.0 & 2.4 \\ 
\end{bmatrix}$ & 5.18 \\
\addlinespace[2ex]
mp-1025678 & \ce{Te6MoW2} & $\begin{bmatrix}
12.6 & 0.0 & 0.0 \\ 
0.0 & 12.6 & 0.0 \\ 
0.0 & 0.0 & 2.4 \\ 
\end{bmatrix}$ & 5.18 \\
\addlinespace[2ex]
mp-1025573 & \ce{Te2W} & $\begin{bmatrix}
12.5 & 0.0 & 0.0 \\ 
0.0 & 12.5 & 0.0 \\ 
0.0 & 0.0 & 2.5 \\ 
\end{bmatrix}$ & 5.06 \\
\addlinespace[2ex]
mp-1030155 & \ce{Te8Mo3W} & $\begin{bmatrix}
14.0 & 0.0 & 0.0 \\ 
0.0 & 14.0 & 0.0 \\ 
0.0 & 0.0 & 2.8 \\ 
\end{bmatrix}$ & 5.03 \\
\addlinespace[2ex]
mp-1030331 & \ce{Te8Mo3W} & $\begin{bmatrix}
13.9 & 0.0 & 0.0 \\ 
0.0 & 13.9 & 0.0 \\ 
0.0 & 0.0 & 2.8 \\ 
\end{bmatrix}$ & 5.02 \\
\addlinespace[2ex]
mp-1023940 & \ce{MoSe2} & $\begin{bmatrix}
9.8 & 0.0 & 0.0 \\ 
0.0 & 9.8 & 0.0 \\ 
0.0 & 0.0 & 2.0 \\ 
\end{bmatrix}$ & 5.01 \\
\addlinespace[2ex]
mp-1030106 & \ce{Te4MoW} & $\begin{bmatrix}
13.7 & 0.0 & 0.0 \\ 
0.0 & 13.7 & 0.0 \\ 
0.0 & 0.0 & 2.8 \\ 
\end{bmatrix}$ & 4.96 \\
\addlinespace[2ex]
mp-1029256 & \ce{Te4MoW} & $\begin{bmatrix}
13.7 & 0.0 & 0.0 \\ 
0.0 & 13.7 & 0.0 \\ 
0.0 & 0.0 & 2.8 \\ 
\end{bmatrix}$ & 4.96 \\
\addlinespace[2ex]
mp-1030335 & \ce{Te4MoW} & $\begin{bmatrix}
13.7 & 0.0 & 0.0 \\ 
0.0 & 13.7 & 0.0 \\ 
0.0 & 0.0 & 2.8 \\ 
\end{bmatrix}$ & 4.95 \\
\addlinespace[2ex]
mp-1028594 & \ce{Te4MoW} & $\begin{bmatrix}
13.7 & 0.0 & 0.0 \\ 
0.0 & 13.7 & 0.0 \\ 
0.0 & 0.0 & 2.8 \\ 
\end{bmatrix}$ & 4.95 \\
\addlinespace[2ex]
mp-1023953 & \ce{MoSeS} & $\begin{bmatrix}
9.6 & 0.0 & 0.0 \\ 
0.0 & 9.6 & 0.0 \\ 
0.0 & 0.0 & 1.9 \\ 
\end{bmatrix}$ & 4.93 \\
\addlinespace[2ex]
mp-1028576 & \ce{Te8MoW3} & $\begin{bmatrix}
13.5 & 0.0 & 0.0 \\ 
0.0 & 13.5 & 0.0 \\ 
0.0 & 0.0 & 2.8 \\ 
\end{bmatrix}$ & 4.88 \\
\addlinespace[2ex]
mp-1018806 & \ce{MoSeS} & $\begin{bmatrix}
10.7 & 0.0 & 0.0 \\ 
0.0 & 10.7 & 0.0 \\ 
0.0 & 0.0 & 2.2 \\ 
\end{bmatrix}$ & 4.84 \\
\addlinespace[2ex]
mp-1030108 & \ce{Te8MoW3} & $\begin{bmatrix}
13.7 & 0.0 & 0.0 \\ 
0.0 & 13.7 & 0.0 \\ 
0.0 & 0.0 & 2.8 \\ 
\end{bmatrix}$ & 4.84 \\
\addlinespace[2ex]
mp-1025799 & \ce{MoSe2} & $\begin{bmatrix}
11.3 & 0.0 & 0.0 \\ 
0.0 & 11.3 & 0.0 \\ 
0.0 & 0.0 & 2.4 \\ 
\end{bmatrix}$ & 4.82 \\
\addlinespace[2ex]
mp-1025906 & \ce{Mo3(Se2S)2} & $\begin{bmatrix}
11.0 & 0.0 & 0.0 \\ 
0.0 & 11.0 & 0.0 \\ 
0.0 & 0.0 & 2.3 \\ 
\end{bmatrix}$ & 4.81 \\
\addlinespace[2ex]
mp-1023928 & \ce{MoWSe4} & $\begin{bmatrix}
9.5 & 0.0 & 0.0 \\ 
0.0 & 9.5 & 0.0 \\ 
0.0 & 0.0 & 2.0 \\ 
\end{bmatrix}$ & 4.81 \\
\addlinespace[2ex]
mp-1025819 & \ce{Mo3(Se2S)2} & $\begin{bmatrix}
11.1 & 0.0 & 0.0 \\ 
0.0 & 11.1 & 0.0 \\ 
0.0 & 0.0 & 2.3 \\ 
\end{bmatrix}$ & 4.78 \\
\addlinespace[2ex]
mp-1025925 & \ce{Mo3(SeS2)2} & $\begin{bmatrix}
10.5 & 0.0 & 0.0 \\ 
0.0 & 10.5 & 0.0 \\ 
0.0 & 0.0 & 2.2 \\ 
\end{bmatrix}$ & 4.76 \\
\addlinespace[2ex]
mp-1023939 & \ce{MoS2} & $\begin{bmatrix}
9.0 & 0.0 & 0.0 \\ 
0.0 & 9.0 & 0.0 \\ 
0.0 & 0.0 & 1.9 \\ 
\end{bmatrix}$ & 4.75 \\
\addlinespace[2ex]
mp-1025988 & \ce{Mo3(SeS2)2} & $\begin{bmatrix}
10.8 & 0.0 & 0.0 \\ 
0.0 & 10.8 & 0.0 \\ 
0.0 & 0.0 & 2.3 \\ 
\end{bmatrix}$ & 4.73 \\
\addlinespace[2ex]
mp-1026023 & \ce{Mo2W(SeS2)2} & $\begin{bmatrix}
10.0 & 0.0 & 0.0 \\ 
0.0 & 10.0 & 0.0 \\ 
0.0 & 0.0 & 2.1 \\ 
\end{bmatrix}$ & 4.68 \\
\addlinespace[2ex]
mp-1025941 & \ce{Mo2W(SeS2)2} & $\begin{bmatrix}
10.1 & 0.0 & 0.0 \\ 
0.0 & 10.1 & 0.0 \\ 
0.0 & 0.0 & 2.2 \\ 
\end{bmatrix}$ & 4.67 \\
\addlinespace[2ex]
mp-1025948 & \ce{Mo2W(SeS2)2} & $\begin{bmatrix}
10.1 & 0.0 & 0.0 \\ 
0.0 & 10.1 & 0.0 \\ 
0.0 & 0.0 & 2.2 \\ 
\end{bmatrix}$ & 4.66 \\
\addlinespace[2ex]
mp-1025874 & \ce{MoS2} & $\begin{bmatrix}
10.2 & 0.0 & 0.0 \\ 
0.0 & 10.2 & 0.0 \\ 
0.0 & 0.0 & 2.2 \\ 
\end{bmatrix}$ & 4.64 \\
\addlinespace[2ex]
mp-1071956 & \ce{TiNF} & $\begin{bmatrix}
8.8 & 0.0 & 0.0 \\ 
0.0 & 9.8 & 0.0 \\ 
0.0 & 0.0 & 2.1 \\ 
\end{bmatrix}$ & 4.64 \\
\addlinespace[2ex]
mp-1023933 & \ce{WSe2} & $\begin{bmatrix}
8.6 & 0.0 & 0.0 \\ 
0.0 & 8.6 & 0.0 \\ 
0.0 & 0.0 & 1.9 \\ 
\end{bmatrix}$ & 4.62 \\
\addlinespace[2ex]
mp-1027492 & \ce{MoSeS} & $\begin{bmatrix}
11.6 & 0.0 & 0.0 \\ 
0.0 & 11.6 & 0.0 \\ 
0.0 & 0.0 & 2.5 \\ 
\end{bmatrix}$ & 4.61 \\
\addlinespace[2ex]
mp-1027687 & \ce{MoSeS} & $\begin{bmatrix}
11.7 & 0.0 & 0.0 \\ 
0.0 & 11.7 & 0.0 \\ 
0.0 & 0.0 & 2.5 \\ 
\end{bmatrix}$ & 4.59 \\
\addlinespace[2ex]
mp-1027795 & \ce{Mo3W(SeS3)2} & $\begin{bmatrix}
10.7 & 0.0 & 0.0 \\ 
0.0 & 10.7 & 0.0 \\ 
0.0 & 0.0 & 2.3 \\ 
\end{bmatrix}$ & 4.58 \\
\addlinespace[2ex]
mp-1027580 & \ce{MoSeS} & $\begin{bmatrix}
11.7 & 0.0 & 0.0 \\ 
0.0 & 11.7 & 0.0 \\ 
0.0 & 0.0 & 2.6 \\ 
\end{bmatrix}$ & 4.58 \\
\addlinespace[2ex]
mp-1026980 & \ce{Mo2Se3S} & $\begin{bmatrix}
12.0 & 0.0 & 0.0 \\ 
0.0 & 12.0 & 0.0 \\ 
0.0 & 0.0 & 2.6 \\ 
\end{bmatrix}$ & 4.57 \\
\addlinespace[2ex]
mp-1026916 & \ce{MoSeS} & $\begin{bmatrix}
11.7 & 0.0 & 0.0 \\ 
0.0 & 11.7 & 0.0 \\ 
0.0 & 0.0 & 2.6 \\ 
\end{bmatrix}$ & 4.57 \\
\addlinespace[2ex]
mp-1027608 & \ce{Mo2SeS3} & $\begin{bmatrix}
11.4 & 0.0 & 0.0 \\ 
0.0 & 11.4 & 0.0 \\ 
0.0 & 0.0 & 2.5 \\ 
\end{bmatrix}$ & 4.55 \\
\addlinespace[2ex]
mp-1025824 & \ce{MoW2(SeS2)2} & $\begin{bmatrix}
9.8 & 0.0 & 0.0 \\ 
0.0 & 9.8 & 0.0 \\ 
0.0 & 0.0 & 2.2 \\ 
\end{bmatrix}$ & 4.54 \\
\addlinespace[2ex]
mp-1027537 & \ce{Mo3W(SeS3)2} & $\begin{bmatrix}
10.9 & 0.0 & 0.0 \\ 
0.0 & 10.9 & 0.0 \\ 
0.0 & 0.0 & 2.4 \\ 
\end{bmatrix}$ & 4.53 \\
\addlinespace[2ex]
mp-1027890 & \ce{Mo2SeS3} & $\begin{bmatrix}
11.5 & 0.0 & 0.0 \\ 
0.0 & 11.5 & 0.0 \\ 
0.0 & 0.0 & 2.5 \\ 
\end{bmatrix}$ & 4.53 \\
\addlinespace[2ex]
mp-1027472 & \ce{Mo3W(SeS3)2} & $\begin{bmatrix}
10.9 & 0.0 & 0.0 \\ 
0.0 & 10.9 & 0.0 \\ 
0.0 & 0.0 & 2.4 \\ 
\end{bmatrix}$ & 4.53 \\
\addlinespace[2ex]
mp-1027646 & \ce{Mo3W(SeS3)2} & $\begin{bmatrix}
10.9 & 0.0 & 0.0 \\ 
0.0 & 10.9 & 0.0 \\ 
0.0 & 0.0 & 2.4 \\ 
\end{bmatrix}$ & 4.53 \\
\addlinespace[2ex]
mp-1027294 & \ce{Mo3W(SeS3)2} & $\begin{bmatrix}
10.9 & 0.0 & 0.0 \\ 
0.0 & 10.9 & 0.0 \\ 
0.0 & 0.0 & 2.4 \\ 
\end{bmatrix}$ & 4.53 \\
\addlinespace[2ex]
mp-1025911 & \ce{Mo2WS6} & $\begin{bmatrix}
9.9 & 0.0 & 0.0 \\ 
0.0 & 9.9 & 0.0 \\ 
0.0 & 0.0 & 2.2 \\ 
\end{bmatrix}$ & 4.52 \\
\addlinespace[2ex]
mp-1025663 & \ce{MoW2(SeS2)2} & $\begin{bmatrix}
9.8 & 0.0 & 0.0 \\ 
0.0 & 9.8 & 0.0 \\ 
0.0 & 0.0 & 2.2 \\ 
\end{bmatrix}$ & 4.52 \\
\addlinespace[2ex]
mp-1023929 & \ce{WSeS} & $\begin{bmatrix}
8.5 & 0.0 & 0.0 \\ 
0.0 & 8.5 & 0.0 \\ 
0.0 & 0.0 & 1.9 \\ 
\end{bmatrix}$ & 4.52 \\
\addlinespace[2ex]
mp-1025599 & \ce{W3(Se2S)2} & $\begin{bmatrix}
9.8 & 0.0 & 0.0 \\ 
0.0 & 9.8 & 0.0 \\ 
0.0 & 0.0 & 2.2 \\ 
\end{bmatrix}$ & 4.47 \\
\addlinespace[2ex]
mp-1025588 & \ce{W3(Se2S)2} & $\begin{bmatrix}
9.8 & 0.0 & 0.0 \\ 
0.0 & 9.8 & 0.0 \\ 
0.0 & 0.0 & 2.2 \\ 
\end{bmatrix}$ & 4.46 \\
\addlinespace[2ex]
mp-754748 & \ce{CoO2} & $\begin{bmatrix}
10.8 & 0.0 & -0.1 \\ 
0.0 & 10.9 & 0.1 \\ 
-0.1 & 0.1 & 2.4 \\ 
\end{bmatrix}$ & 4.45 \\
\addlinespace[2ex]
mp-1027274 & \ce{MoWSeS3} & $\begin{bmatrix}
10.6 & 0.0 & 0.0 \\ 
0.0 & 10.6 & 0.0 \\ 
0.0 & 0.0 & 2.4 \\ 
\end{bmatrix}$ & 4.44 \\
\addlinespace[2ex]
mp-1030745 & \ce{MoWSeS3} & $\begin{bmatrix}
10.6 & 0.0 & 0.0 \\ 
0.0 & 10.6 & 0.0 \\ 
0.0 & 0.0 & 2.4 \\ 
\end{bmatrix}$ & 4.44 \\
\addlinespace[2ex]
mp-1027292 & \ce{MoWSeS3} & $\begin{bmatrix}
10.6 & 0.0 & 0.0 \\ 
0.0 & 10.6 & 0.0 \\ 
0.0 & 0.0 & 2.4 \\ 
\end{bmatrix}$ & 4.44 \\
\addlinespace[2ex]
mp-1026975 & \ce{MoWSeS3} & $\begin{bmatrix}
10.6 & 0.0 & 0.0 \\ 
0.0 & 10.6 & 0.0 \\ 
0.0 & 0.0 & 2.4 \\ 
\end{bmatrix}$ & 4.43 \\
\addlinespace[2ex]
mp-1027391 & \ce{MoWSeS3} & $\begin{bmatrix}
10.6 & 0.0 & 0.0 \\ 
0.0 & 10.6 & 0.0 \\ 
0.0 & 0.0 & 2.4 \\ 
\end{bmatrix}$ & 4.43 \\
\addlinespace[2ex]
mp-1030146 & \ce{MoWSeS3} & $\begin{bmatrix}
10.6 & 0.0 & 0.0 \\ 
0.0 & 10.6 & 0.0 \\ 
0.0 & 0.0 & 2.4 \\ 
\end{bmatrix}$ & 4.43 \\
\addlinespace[2ex]
mp-1026034 & \ce{Mo(WS3)2} & $\begin{bmatrix}
9.6 & 0.0 & 0.0 \\ 
0.0 & 9.6 & 0.0 \\ 
0.0 & 0.0 & 2.2 \\ 
\end{bmatrix}$ & 4.40 \\
\addlinespace[2ex]
mp-1027159 & \ce{MoW(SeS)2} & $\begin{bmatrix}
11.2 & 0.0 & 0.0 \\ 
0.0 & 11.2 & 0.0 \\ 
0.0 & 0.0 & 2.6 \\ 
\end{bmatrix}$ & 4.39 \\
\addlinespace[2ex]
mp-1025689 & \ce{Mo(WS3)2} & $\begin{bmatrix}
9.7 & 0.0 & 0.0 \\ 
0.0 & 9.7 & 0.0 \\ 
0.0 & 0.0 & 2.2 \\ 
\end{bmatrix}$ & 4.39 \\
\addlinespace[2ex]
mp-1025584 & \ce{W3(SeS2)2} & $\begin{bmatrix}
9.6 & 0.0 & 0.0 \\ 
0.0 & 9.6 & 0.0 \\ 
0.0 & 0.0 & 2.2 \\ 
\end{bmatrix}$ & 4.39 \\
\addlinespace[2ex]
mp-1027569 & \ce{Mo3WS8} & $\begin{bmatrix}
10.8 & 0.0 & 0.0 \\ 
0.0 & 10.8 & 0.0 \\ 
0.0 & 0.0 & 2.5 \\ 
\end{bmatrix}$ & 4.39 \\
\addlinespace[2ex]
mp-1027645 & \ce{Mo3WS8} & $\begin{bmatrix}
10.8 & 0.0 & 0.0 \\ 
0.0 & 10.8 & 0.0 \\ 
0.0 & 0.0 & 2.5 \\ 
\end{bmatrix}$ & 4.38 \\
\addlinespace[2ex]
mp-1025577 & \ce{W3(SeS2)2} & $\begin{bmatrix}
9.8 & 0.0 & 0.0 \\ 
0.0 & 9.8 & 0.0 \\ 
0.0 & 0.0 & 2.3 \\ 
\end{bmatrix}$ & 4.36 \\
\addlinespace[2ex]
mp-1030520 & \ce{MoW3(SeS3)2} & $\begin{bmatrix}
10.4 & 0.0 & 0.0 \\ 
0.0 & 10.4 & 0.0 \\ 
0.0 & 0.0 & 2.4 \\ 
\end{bmatrix}$ & 4.35 \\
\addlinespace[2ex]
mp-1023925 & \ce{WS2} & $\begin{bmatrix}
7.7 & 0.0 & 0.0 \\ 
0.0 & 7.7 & 0.0 \\ 
0.0 & 0.0 & 1.8 \\ 
\end{bmatrix}$ & 4.35 \\
\addlinespace[2ex]
mp-1029037 & \ce{MoW3(SeS3)2} & $\begin{bmatrix}
10.4 & 0.0 & 0.0 \\ 
0.0 & 10.4 & 0.0 \\ 
0.0 & 0.0 & 2.4 \\ 
\end{bmatrix}$ & 4.34 \\
\addlinespace[2ex]
mp-1028769 & \ce{W2Se3S} & $\begin{bmatrix}
10.6 & 0.0 & 0.0 \\ 
0.0 & 10.6 & 0.0 \\ 
0.0 & 0.0 & 2.4 \\ 
\end{bmatrix}$ & 4.33 \\
\addlinespace[2ex]
mp-1028686 & \ce{W2Se3S} & $\begin{bmatrix}
10.6 & 0.0 & 0.0 \\ 
0.0 & 10.6 & 0.0 \\ 
0.0 & 0.0 & 2.4 \\ 
\end{bmatrix}$ & 4.32 \\
\addlinespace[2ex]
mp-1030566 & \ce{MoW3(SeS)4} & $\begin{bmatrix}
10.9 & 0.0 & 0.0 \\ 
0.0 & 10.9 & 0.0 \\ 
0.0 & 0.0 & 2.5 \\ 
\end{bmatrix}$ & 4.32 \\
\addlinespace[2ex]
mp-1027269 & \ce{MoWS4} & $\begin{bmatrix}
10.5 & 0.0 & 0.0 \\ 
0.0 & 10.5 & 0.0 \\ 
0.0 & 0.0 & 2.4 \\ 
\end{bmatrix}$ & 4.31 \\
\addlinespace[2ex]
mp-1028488 & \ce{WSeS} & $\begin{bmatrix}
10.4 & 0.0 & 0.0 \\ 
0.0 & 10.4 & 0.0 \\ 
0.0 & 0.0 & 2.4 \\ 
\end{bmatrix}$ & 4.31 \\
\addlinespace[2ex]
mp-1028772 & \ce{WSeS} & $\begin{bmatrix}
10.4 & 0.0 & 0.0 \\ 
0.0 & 10.4 & 0.0 \\ 
0.0 & 0.0 & 2.4 \\ 
\end{bmatrix}$ & 4.30 \\
\addlinespace[2ex]
mp-1028663 & \ce{WSeS} & $\begin{bmatrix}
10.4 & 0.0 & 0.0 \\ 
0.0 & 10.4 & 0.0 \\ 
0.0 & 0.0 & 2.4 \\ 
\end{bmatrix}$ & 4.30 \\
\addlinespace[2ex]
mp-1027647 & \ce{MoWS4} & $\begin{bmatrix}
10.5 & 0.0 & 0.0 \\ 
0.0 & 10.5 & 0.0 \\ 
0.0 & 0.0 & 2.4 \\ 
\end{bmatrix}$ & 4.30 \\
\addlinespace[2ex]
mp-1028764 & \ce{WSeS} & $\begin{bmatrix}
10.4 & 0.0 & 0.0 \\ 
0.0 & 10.4 & 0.0 \\ 
0.0 & 0.0 & 2.4 \\ 
\end{bmatrix}$ & 4.29 \\
\addlinespace[2ex]
mp-1030119 & \ce{MoWS4} & $\begin{bmatrix}
10.6 & 0.0 & 0.0 \\ 
0.0 & 10.6 & 0.0 \\ 
0.0 & 0.0 & 2.5 \\ 
\end{bmatrix}$ & 4.29 \\
\addlinespace[2ex]
mp-1025571 & \ce{WS2} & $\begin{bmatrix}
9.3 & 0.0 & 0.0 \\ 
0.0 & 9.3 & 0.0 \\ 
0.0 & 0.0 & 2.2 \\ 
\end{bmatrix}$ & 4.29 \\
\addlinespace[2ex]
mp-1027273 & \ce{MoW3S8} & $\begin{bmatrix}
10.3 & 0.0 & 0.0 \\ 
0.0 & 10.3 & 0.0 \\ 
0.0 & 0.0 & 2.4 \\ 
\end{bmatrix}$ & 4.23 \\
\addlinespace[2ex]
mp-1029246 & \ce{MoW3S8} & $\begin{bmatrix}
10.3 & 0.0 & 0.0 \\ 
0.0 & 10.3 & 0.0 \\ 
0.0 & 0.0 & 2.5 \\ 
\end{bmatrix}$ & 4.21 \\
\addlinespace[2ex]
mp-1028558 & \ce{W2SeS3} & $\begin{bmatrix}
10.4 & 0.0 & 0.0 \\ 
0.0 & 10.4 & 0.0 \\ 
0.0 & 0.0 & 2.5 \\ 
\end{bmatrix}$ & 4.19 \\
\addlinespace[2ex]
mp-1278455 & \ce{CoO2} & $\begin{bmatrix}
12.3 & -0.1 & -2.9 \\ 
-0.1 & 12.4 & -1.6 \\ 
-2.9 & -1.6 & 4.4 \\ 
\end{bmatrix}$ & 4.17 \\
\addlinespace[2ex]
mp-1028441 & \ce{WS2} & $\begin{bmatrix}
10.0 & 0.0 & 0.0 \\ 
0.0 & 10.0 & 0.0 \\ 
0.0 & 0.0 & 2.4 \\ 
\end{bmatrix}$ & 4.16 \\
\addlinespace[2ex]
mp-1227328 & \ce{Ca(BC3)2} & $\begin{bmatrix}
14.8 & 0.0 & 0.0 \\ 
0.0 & 3.7 & 0.0 \\ 
0.0 & 0.0 & 13.4 \\ 
\end{bmatrix}$ & 4.04 \\
\addlinespace[2ex]
mp-1318786 & \ce{LiCo3NiO8} & $\begin{bmatrix}
12.6 & 0.4 & 0.0 \\ 
0.4 & 12.3 & 0.0 \\ 
0.0 & 0.0 & 3.2 \\ 
\end{bmatrix}$ & 4.04 \\
\addlinespace[2ex]
mp-1411545 & \ce{NiS2} & $\begin{bmatrix}
7.1 & 0.0 & -2.7 \\ 
0.0 & 26.6 & 0.0 \\ 
-2.7 & 0.0 & 26.2 \\ 
\end{bmatrix}$ & 3.97 \\
\addlinespace[2ex]
mp-1272680 & \ce{CoO2} & $\begin{bmatrix}
11.9 & 0.0 & 0.1 \\ 
0.0 & 11.9 & -0.2 \\ 
0.1 & -0.2 & 3.0 \\ 
\end{bmatrix}$ & 3.97 \\
\addlinespace[2ex]
mp-752738 & \ce{Co3NiO8} & $\begin{bmatrix}
4.0 & -2.4 & -1.5 \\ 
-2.4 & 11.2 & -0.4 \\ 
-1.5 & -0.4 & 11.7 \\ 
\end{bmatrix}$ & 3.96 \\
\addlinespace[2ex]
mp-773511 & \ce{Co5NiO12} & $\begin{bmatrix}
12.1 & -0.1 & 0.0 \\ 
-0.1 & 12.0 & 0.0 \\ 
0.0 & 0.0 & 3.1 \\ 
\end{bmatrix}$ & 3.88 \\
\addlinespace[2ex]
mp-1314136 & \ce{LiMnCo3O8} & $\begin{bmatrix}
5.7 & -0.5 & 3.6 \\ 
-0.5 & 11.2 & -0.8 \\ 
3.6 & -0.8 & 8.5 \\ 
\end{bmatrix}$ & 3.72 \\
\addlinespace[2ex]
mp-1066781 & \ce{BrCl} & $\begin{bmatrix}
2.7 & 0.0 & 0.0 \\ 
0.0 & 1.9 & 0.0 \\ 
0.0 & 0.0 & 7.1 \\ 
\end{bmatrix}$ & 3.65 \\
\addlinespace[2ex]
mp-763057 & \ce{Mn(CoO3)2} & $\begin{bmatrix}
11.2 & 0.0 & 0.0 \\ 
0.0 & 6.6 & -4.0 \\ 
0.0 & -4.0 & 7.7 \\ 
\end{bmatrix}$ & 3.59 \\
\addlinespace[2ex]
mp-27213 & \ce{AuBr3} & $\begin{bmatrix}
8.5 & 0.0 & 0.0 \\ 
0.0 & 2.9 & 0.0 \\ 
0.0 & 0.0 & 9.9 \\ 
\end{bmatrix}$ & 3.39 \\
\addlinespace[2ex]
mp-1296423 & \ce{LiMnCo3O8} & $\begin{bmatrix}
10.1 & 0.2 & 0.0 \\ 
0.2 & 10.2 & 0.0 \\ 
0.0 & 0.0 & 3.1 \\ 
\end{bmatrix}$ & 3.38 \\
\addlinespace[2ex]
mp-1285961 & \ce{Li(CoO2)3} & $\begin{bmatrix}
10.5 & -0.2 & -0.1 \\ 
-0.2 & 10.2 & 0.1 \\ 
-0.1 & 0.1 & 3.2 \\ 
\end{bmatrix}$ & 3.36 \\
\addlinespace[2ex]
mp-1303340 & \ce{LiMnCo3O8} & $\begin{bmatrix}
5.2 & -0.1 & 3.2 \\ 
-0.1 & 10.1 & 0.0 \\ 
3.2 & 0.0 & 7.7 \\ 
\end{bmatrix}$ & 3.34 \\
\addlinespace[2ex]
mp-755555 & \ce{Mn5CoO12} & $\begin{bmatrix}
9.8 & -0.1 & 0.0 \\ 
-0.1 & 9.8 & 0.0 \\ 
0.0 & 0.0 & 2.9 \\ 
\end{bmatrix}$ & 3.34 \\
\addlinespace[2ex]
mp-755862 & \ce{LiMnCo3O8} & $\begin{bmatrix}
5.1 & -0.1 & 3.1 \\ 
-0.1 & 9.5 & -0.1 \\ 
3.1 & -0.1 & 7.6 \\ 
\end{bmatrix}$ & 3.22 \\
\addlinespace[2ex]
mp-1184859 & \ce{HI3} & $\begin{bmatrix}
3.0 & -1.8 & 0.0 \\ 
-1.8 & 4.6 & 0.0 \\ 
0.0 & 0.0 & 1.8 \\ 
\end{bmatrix}$ & 3.21 \\
\addlinespace[2ex]
mp-759301 & \ce{Li(CoO2)3} & $\begin{bmatrix}
10.6 & 0.1 & 0.0 \\ 
0.1 & 10.7 & 0.0 \\ 
0.0 & 0.0 & 3.4 \\ 
\end{bmatrix}$ & 3.12 \\
\addlinespace[2ex]
mp-759163 & \ce{VOF2} & $\begin{bmatrix}
3.2 & 0.0 & 0.0 \\ 
0.0 & 7.9 & -3.2 \\ 
0.0 & -3.2 & 5.5 \\ 
\end{bmatrix}$ & 3.12 \\
\addlinespace[2ex]
mp-1217314 & \ce{TeMo2Se3} & $\begin{bmatrix}
16.4 & 0.0 & 0.0 \\ 
0.0 & 16.4 & 0.0 \\ 
0.0 & 0.0 & 5.3 \\ 
\end{bmatrix}$ & 3.09 \\
\addlinespace[2ex]
mp-1018809 & \ce{MoS2} & $\begin{bmatrix}
14.3 & 0.0 & 0.0 \\ 
0.0 & 14.3 & 0.0 \\ 
0.0 & 0.0 & 4.7 \\ 
\end{bmatrix}$ & 3.03 \\
\addlinespace[2ex]
mp-1221404 & \ce{MoSeS} & $\begin{bmatrix}
15.0 & 0.0 & 0.0 \\ 
0.0 & 15.0 & 0.0 \\ 
0.0 & 0.0 & 5.0 \\ 
\end{bmatrix}$ & 3.02 \\
\addlinespace[2ex]
mp-9481 & \ce{TcS2} & $\begin{bmatrix}
15.6 & -1.2 & 2.0 \\ 
-1.2 & 14.2 & 3.8 \\ 
2.0 & 3.8 & 7.7 \\ 
\end{bmatrix}$ & 2.99 \\
\addlinespace[2ex]
mp-572758 & \ce{ReS2} & $\begin{bmatrix}
14.0 & -0.2 & 0.0 \\ 
-0.2 & 12.7 & 2.0 \\ 
0.0 & 2.0 & 5.3 \\ 
\end{bmatrix}$ & 2.96 \\
\addlinespace[2ex]
mp-1217371 & \ce{Te3Mo2Se} & $\begin{bmatrix}
17.9 & 0.0 & 0.0 \\ 
0.0 & 17.9 & 0.0 \\ 
0.0 & 0.0 & 6.0 \\ 
\end{bmatrix}$ & 2.96 \\
\addlinespace[2ex]
mp-754774 & \ce{Li2Mn3(CoO4)3} & $\begin{bmatrix}
8.9 & 0.1 & -0.1 \\ 
0.1 & 4.8 & -2.6 \\ 
-0.1 & -2.6 & 7.1 \\ 
\end{bmatrix}$ & 2.93 \\
\addlinespace[2ex]
mp-1018807 & \ce{MoSe2} & $\begin{bmatrix}
15.7 & 0.0 & 0.0 \\ 
0.0 & 15.7 & 0.0 \\ 
0.0 & 0.0 & 5.5 \\ 
\end{bmatrix}$ & 2.86 \\
\addlinespace[2ex]
mp-1219546 & \ce{ReSeS} & $\begin{bmatrix}
6.1 & -2.8 & 0.0 \\ 
-2.8 & 13.0 & -0.5 \\ 
0.0 & -0.5 & 14.3 \\ 
\end{bmatrix}$ & 2.84 \\
\addlinespace[2ex]
mp-1221485 & \ce{Mo2SeS3} & $\begin{bmatrix}
15.0 & 0.0 & 0.0 \\ 
0.0 & 15.0 & 0.0 \\ 
0.0 & 0.0 & 5.4 \\ 
\end{bmatrix}$ & 2.80 \\
\addlinespace[2ex]
mp-753228 & \ce{LiMn3O6} & $\begin{bmatrix}
8.2 & 0.0 & 0.0 \\ 
0.0 & 7.6 & -0.1 \\ 
0.0 & -0.1 & 3.1 \\ 
\end{bmatrix}$ & 2.62 \\
\addlinespace[2ex]
mp-1276496 & \ce{NaV2O4} & $\begin{bmatrix}
2.7 & 0.0 & 0.1 \\ 
0.0 & 5.4 & 0.1 \\ 
0.1 & 0.1 & 7.0 \\ 
\end{bmatrix}$ & 2.62 \\
\addlinespace[2ex]
mp-861871 & \ce{SeI2} & $\begin{bmatrix}
5.0 & 0.0 & 0.0 \\ 
0.0 & 5.0 & 0.0 \\ 
0.0 & 0.0 & 13.0 \\ 
\end{bmatrix}$ & 2.58 \\
\addlinespace[2ex]
mp-752885 & \ce{LiVOF3} & $\begin{bmatrix}
3.2 & -0.8 & -0.9 \\ 
-0.8 & 4.6 & 1.8 \\ 
-0.9 & 1.8 & 4.9 \\ 
\end{bmatrix}$ & 2.53 \\
\addlinespace[2ex]
mp-756552 & \ce{Mg(NiO2)4} & $\begin{bmatrix}
8.9 & 0.0 & 0.0 \\ 
0.0 & 8.9 & 0.0 \\ 
0.0 & 0.0 & 3.6 \\ 
\end{bmatrix}$ & 2.51 \\
\addlinespace[2ex]
mp-1016190 & \ce{KMn3O6} & $\begin{bmatrix}
6.0 & 0.0 & 0.0 \\ 
0.0 & 2.4 & 0.0 \\ 
0.0 & 0.0 & 5.5 \\ 
\end{bmatrix}$ & 2.49 \\
\addlinespace[2ex]
mp-1273655 & \ce{MnOF} & $\begin{bmatrix}
4.0 & -0.6 & 0.1 \\ 
-0.6 & 6.0 & 0.9 \\ 
0.1 & 0.9 & 2.9 \\ 
\end{bmatrix}$ & 2.43 \\
\addlinespace[2ex]
mp-1223545 & \ce{KMn2O4} & $\begin{bmatrix}
2.8 & 0.0 & 0.2 \\ 
0.0 & 6.7 & 0.0 \\ 
0.2 & 0.0 & 5.9 \\ 
\end{bmatrix}$ & 2.42 \\
\addlinespace[2ex]
mp-997108 & \ce{RbAgO2} & $\begin{bmatrix}
4.6 & -1.4 & -0.6 \\ 
-1.4 & 4.0 & 0.1 \\ 
-0.6 & 0.1 & 2.7 \\ 
\end{bmatrix}$ & 2.42 \\
\addlinespace[2ex]
mp-2422143 & \ce{Pd(Se3Br)2} & $\begin{bmatrix}
3.8 & 0.0 & -0.2 \\ 
0.0 & 6.9 & -1.5 \\ 
-0.2 & -1.5 & 8.3 \\ 
\end{bmatrix}$ & 2.42 \\
\addlinespace[2ex]
mp-1104174 & \ce{K4(NiO2)3} & $\begin{bmatrix}
2.7 & 0.0 & 0.4 \\ 
0.0 & 3.4 & 0.0 \\ 
0.4 & 0.0 & 1.6 \\ 
\end{bmatrix}$ & 2.41 \\
\addlinespace[2ex]
mp-1101129 & \ce{ThPbI12} & $\begin{bmatrix}
7.6 & 0.0 & 0.0 \\ 
0.0 & 4.1 & 0.0 \\ 
0.0 & 0.0 & 3.2 \\ 
\end{bmatrix}$ & 2.37 \\
\addlinespace[2ex]
mp-1018888 & \ce{PdCl2} & $\begin{bmatrix}
5.6 & 0.0 & 0.0 \\ 
0.0 & 2.4 & 0.2 \\ 
0.0 & 0.2 & 3.4 \\ 
\end{bmatrix}$ & 2.33 \\
\addlinespace[2ex]
mp-1003484 & \ce{MgMn4O8} & $\begin{bmatrix}
6.7 & 0.2 & 1.0 \\ 
0.2 & 7.7 & -0.8 \\ 
1.0 & -0.8 & 3.9 \\ 
\end{bmatrix}$ & 2.31 \\
\addlinespace[2ex]
mp-758725 & \ce{Li2(CoO2)3} & $\begin{bmatrix}
9.1 & 0.0 & 0.0 \\ 
0.0 & 9.1 & 0.0 \\ 
0.0 & 0.0 & 4.0 \\ 
\end{bmatrix}$ & 2.31 \\
\addlinespace[2ex]
mp-1068977 & \ce{K2PdC2} & $\begin{bmatrix}
3.6 & 0.0 & 0.0 \\ 
0.0 & 3.6 & 0.0 \\ 
0.0 & 0.0 & 8.1 \\ 
\end{bmatrix}$ & 2.29 \\
\addlinespace[2ex]
mp-20343 & \ce{NaAuO2} & $\begin{bmatrix}
2.8 & 0.0 & 0.0 \\ 
0.0 & 3.5 & 0.0 \\ 
0.0 & 0.0 & 6.2 \\ 
\end{bmatrix}$ & 2.25 \\
\addlinespace[2ex]
mp-3342 & \ce{ZnPS3} & $\begin{bmatrix}
6.8 & 0.0 & 0.0 \\ 
0.0 & 6.8 & 0.0 \\ 
0.0 & 0.0 & 3.3 \\ 
\end{bmatrix}$ & 2.04 \\
\addlinespace[2ex]
mp-772788 & \ce{Ba2Cu2O5} & $\begin{bmatrix}
4.1 & 0.0 & 0.0 \\ 
0.0 & 8.0 & 0.0 \\ 
0.0 & 0.0 & 5.8 \\ 
\end{bmatrix}$ & 1.97 \\
\addlinespace[2ex]
mp-1193708 & \ce{Al2(PS3)3} & $\begin{bmatrix}
6.5 & 0.0 & 0.0 \\ 
0.0 & 6.5 & 0.0 \\ 
0.0 & 0.0 & 3.5 \\ 
\end{bmatrix}$ & 1.88 \\
\addlinespace[2ex]
mp-569017 & \ce{PdI2} & $\begin{bmatrix}
12.8 & 0.0 & 0.0 \\ 
0.0 & 7.0 & 0.0 \\ 
0.0 & 0.0 & 7.8 \\ 
\end{bmatrix}$ & 1.82 \\
\addlinespace[2ex]
mp-580886 & \ce{ZrIN} & $\begin{bmatrix}
5.6 & 0.0 & -0.2 \\ 
0.0 & 7.9 & 0.0 \\ 
-0.2 & 0.0 & 7.9 \\ 
\end{bmatrix}$ & 1.42 \\
\addlinespace[2ex]
mp-567441 & \ce{HfIN} & $\begin{bmatrix}
5.4 & 0.0 & -0.2 \\ 
0.0 & 7.2 & 0.0 \\ 
-0.2 & 0.0 & 7.2 \\ 
\end{bmatrix}$ & 1.34 \\
\addlinespace[2ex]
mp-561973 & \ce{K(TeO3)2} & $\begin{bmatrix}
4.4 & 0.0 & 0.0 \\ 
0.0 & 4.0 & 0.2 \\ 
0.0 & 0.2 & 4.5 \\ 
\end{bmatrix}$ & 1.16 \\
\bottomrule
\end{longtable}

\bibliography{refs}